\documentclass[10pt,journal,compsoc]{IEEEtran}
\ifCLASSOPTIONcompsoc
  \usepackage[nocompress]{cite}
\else
  \usepackage{cite}
\fi
\usepackage{times}
\usepackage{helvet}
\usepackage{courier}
\usepackage{latexsym,bm,amsmath,amssymb,epstopdf}
\usepackage{algorithmic}
\usepackage{cite}
\usepackage{graphicx}
\usepackage{subfigure}

\usepackage{colortbl}
\usepackage{array}
\usepackage{caption}
\usepackage{makecell}
\usepackage{multirow}
\usepackage{threeparttable}
\usepackage{amsmath}
\usepackage{booktabs}
\usepackage[ruled]{algorithm2e}

\usepackage[normalem]{ulem}
\graphicspath{
{./figs/}
}

\begin{document}
\title{Shielding Collaborative Learning: Mitigating Poisoning Attacks through Client-Side Detection}

\author{Lingchen~Zhao,
        Shengshan~Hu,
        Qian~Wang,~\IEEEmembership{Senior Member,~IEEE,}
        Jianlin~Jiang, \\
        Chao~Shen,
        Xiangyang~Luo, and~Pengfei~Hu

\IEEEcompsocitemizethanks{
\IEEEcompsocthanksitem L. Zhao and Q. Wang are with the Key Laboratory of Aerospace Information Security and Trusted Computing, Ministry of Education, the School of Cyber Science and Engineering, Wuhan University, Wuhan 430072, Hubei, China, email: \{lczhaocs, qianwang\}@whu.edu.cn.
\IEEEcompsocthanksitem S. Hu is with Hubei Engineering Research Center on Big Data Security, School of Cyber Science and Engineering, Huazhong University of Science and Technology, Wuhan 430074, Hubei,  China, email: hushengshan@hust.edu.cn.
\IEEEcompsocthanksitem J. Jiang is with the School of Computer Science, Wuhan University, Wuhan 430072, Hubei, China, email: jianlinjiang@whu.edu.cn.
\IEEEcompsocthanksitem C. Shen is with the MOE Key Laboratory for Intelligent Networks and Network Security, and also with the School of Cyber Science and Engineering, Xi'an Jiaotong University, Xi'an 710049, Shaanxi, China, email: chaoshen@mail.xjtu.edu.cn.
\IEEEcompsocthanksitem X. Luo is with the State Key Laboratory of Mathematical Engineering and Advanced Computing, Zhengzhou 450002, Henan, China, email: xiangyangluo@126.com.
\IEEEcompsocthanksitem P. Hu is with the VMware, Inc., 3401 Hillview Ave, Palo Alto 94304, CA, USA, email: pengfeih@vmware.com.
}}

\markboth{Journal of \LaTeX\ Class Files,~Vol.~14, No.~8, August~2015}%
{Shell \MakeLowercase{\textit{et al.}}: Bare Demo of IEEEtran.cls for Computer Society Journals}

\IEEEtitleabstractindextext{%
\begin{abstract}
Collaborative learning allows multiple clients to train a joint model without sharing their data with each other. Each client performs training locally and then submits the model updates to a central server for aggregation. Since the server has no visibility into the process of generating the updates, collaborative learning is vulnerable to poisoning attacks where a malicious client can generate a poisoned update to introduce backdoor functionality to the joint model. The existing solutions for detecting poisoned updates, however, fail to defend against the recently proposed attacks, especially in the non-IID (independent and identically distributed) setting.
In this paper, we present a novel defense scheme to detect anomalous updates in both IID and non-IID settings.
Our key idea is to realize client-side cross-validation, where each update is evaluated over other clients' local data. The server will adjust the weights of the updates based on the evaluation results when performing aggregation.
To adapt to the unbalanced distribution of data in the non-IID setting, a dynamic client allocation mechanism is designed to assign detection tasks to the most suitable clients.
During the detection process, we also protect the client-level privacy to prevent malicious clients from knowing the participations of other clients, by integrating differential privacy with our design without degrading the detection performance.
Our experimental evaluations on three real-world datasets show that our scheme is significantly robust to two representative poisoning attacks.
\end{abstract}

}

\maketitle

\IEEEdisplaynontitleabstractindextext

\IEEEpeerreviewmaketitle

\IEEEraisesectionheading{\section{Introduction}\label{sec:introduction}}
\IEEEPARstart{C}{ollaborative} learning is an attractive framework for implementing distributed learning with massive clients. In contrast to conventional machine learning approaches which require a central data center, collaborative learning allows clients to perform training locally, and only share the local model rather than the training data.
In each training round, a subset of clients are selected, each of which downloads the current global model and computes an updated model based on their local data. The model updates are sent to the server, who is responsible for aggregation, to construct an improved global model. Motivating examples include training next-word predictors or speech classifiers on large-scale users' smartphones.

\begin{figure}[t!]
  \centering
  \includegraphics[width=0.95\columnwidth]{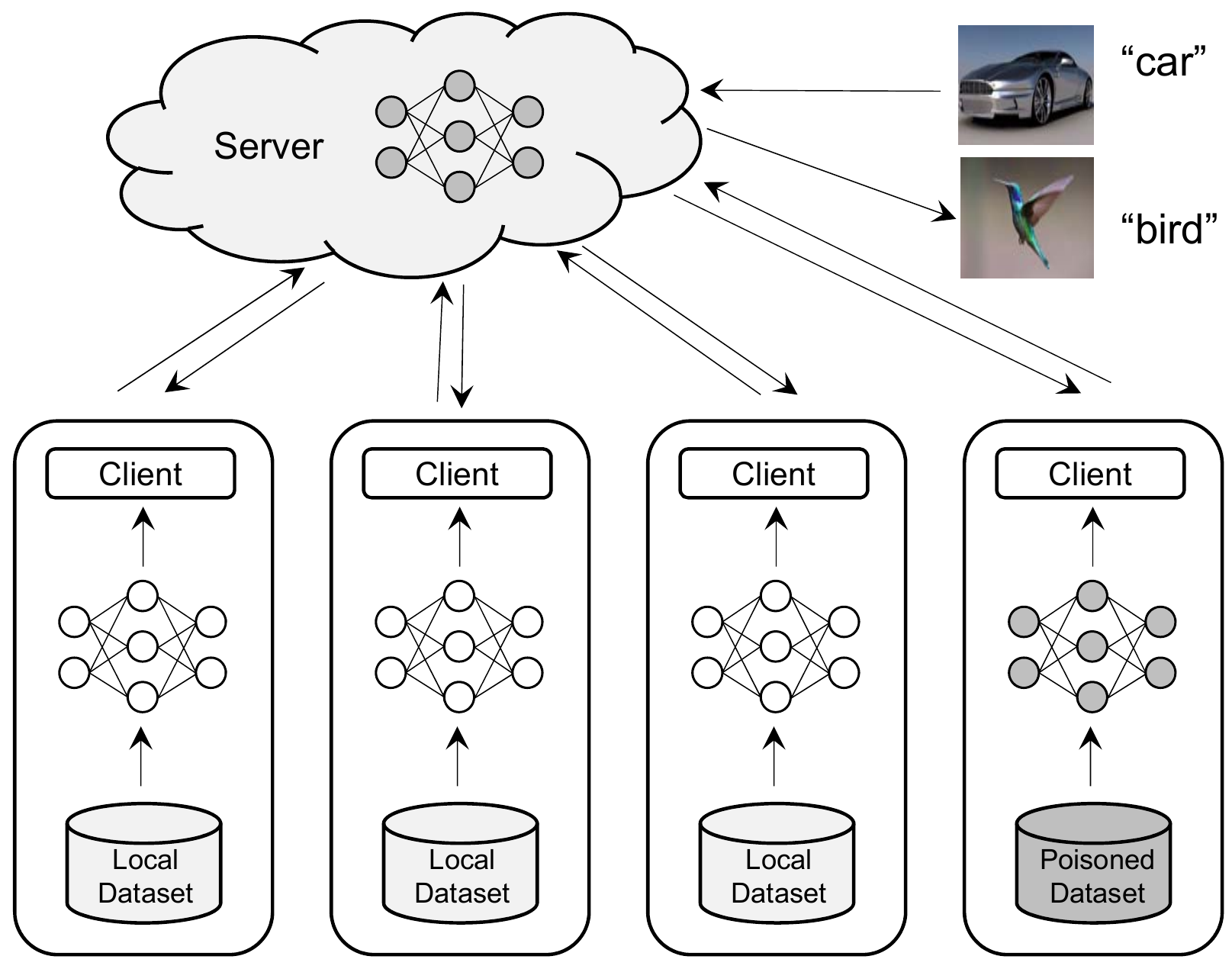}
  %\vspace{-2mm}
  \caption{Overview of the poisoning attack}\label{fig:model}
  %\vspace{-4mm}
\end{figure}

To make full use of clients' data without revealing their privacy, the server by design has no visibility into the local data and training process  in collaborative learning. This also opens the door to poisoning attacks where malicious clients can create poisoned updates to introduce backdoor functionality to the global model\cite{shen2016uror, fung2018mitigating}. In particular, recent works show that a poisoned update can control the behavior of the global model on an attacker-chosen backdoor subtask while maintaining a good performance on the collaborated learning task~\cite{bagdasaryan2018backdoor}. Figure~\ref{fig:model} presents an overview of the poisoning attack.

In centralized learning where the server has access to the training data, detecting poisoned models can be realized by directly evaluating the trained model on data samples~\cite{baracaldo2017mitigating}, or utilizing another trusted dataset to retrain a detector~\cite{steinhardt2017certified, kloft2012security}. In collaborative learning, however, these methods are not applicable since the server can no longer access the training data. More importantly, the poisoned model can still maintain a high accuracy on the main task, making it more difficult to detect anomalies. To solve this problem, solutions based on statistical analysis~\cite{chen2017distributed, yin2018byzantine} may be the potential countermeasure. Auror~\cite{shen2016uror} showed that if the malicious client continually submits poisoned models in every round, misbehaviors can be found by measuring the difference of distributions between benign and malicious updates through clustering. Krum~\cite{blanchard2017machine} and its variants~\cite{chen2017distributed, yin2018byzantine} tried to improve the robustness of the model by removing updates that are far from the average. Nevertheless, both of these two methods are not suitable for non-IID training data~\cite{bagdasaryan2018backdoor}. Recently FoolsGold~\cite{fung2018mitigating} is designed to detect the sybil attack, which identifies malicious clients by calculating the similarities between different updates. Yet, it can be evaded by performing a single-client attack or decomposing the poisoned model into several orthogonal vectors.

In this paper, we design a new method to effectively detect anomalous updates in collaborative learning. Instead of focusing on equipping the server, we propose delegating the detection tasks to the clients whose private data can be used to evaluate the performance of the updates. Based on the evaluation results, the server can then adjust the weights  when aggregating the updates. Our method remains effective not only for the IID setting, but also for the non-IID setting. During the detection process, we also aim to prevent malicious clients from knowing the participation of other clients. We integrate differential privacy to the delegated detection tasks to protect the client-level privacy without degrading the detection performance. Our experimental evaluations on the MNIST, KDDCup and CIFAR-10 datasets show that our client-side cross-validation is significantly robust to two representative poisoning attacks: the label-flipping attack and the backdoor attack.
In summary, we offer the following contributions.

\begin{itemize}
	\item We present a novel scheme to defend against poisoning attacks in collaborative learning by delegating detection tasks to clients for cross-validation, such that anomalous updates can be easily detected in the IID setting.

	\item We extend our cross-validation architecture to the non-IID setting. A dynamic client allocation method is designed to assign the detection task to the most suitable clients. Besides, we protect the client-level privacy to prevent the potential leakages about honest clients' datasets.

	\item We evaluate our scheme experimentally on two real-world datasets against two representative poisoning attacks. The results demonstrate that our scheme can detect poisoned updates with high probability, and cause a negligible impact on the accuracy of the global model.
\end{itemize}

\section{Background}
In this section, we provide necessary background knowledge about machine learning, collaborative learning and poisoning attacks.
\subsection{Machine Learning}
In this paper, we focus on the generic setting of supervised learning for classification tasks. Let $\mathcal{D}$ be a dataset that contains $n$ samples $\mathbf{x}_1, \mathbf{x}_2, \ldots, \mathbf{x}_n$, each of which sample has $d$ features and corresponds to a label $y_i$. The goal of a machine learning algorithm is to obtain a model $\mathcal{M}$ that takes a sample $\mathbf{x}$ as input and outputs a predicted value $y'$ which should be as close as possible to its ground truth $y$. In general, $\mathbf{x}$ can be parameterized with a vector or a matrix which consists of real numbers. In order to find the optimal set of parameters for $\mathcal{M}$, the training algorithm tries to minimize a loss function $\mathcal{Q}(\mathbf{w})=\frac{1}{n}\sum_{i=1}^{n}\mathcal{Q}_i(\mathbf{w})$, where $\mathbf{w}$ is the vector/matrix of the model parameters, and $\mathcal{Q}_i$ is the loss value for the $i$-th sample in $\mathcal{D}$.

\textbf{Stochastic Gradient Descent (SGD).} SGD is the most common method to minimize the loss function,  which iteratively seeks for the minimum value in each dimension, i.e., $\mathbf{w}=\mathbf{w}-\frac{\alpha}{n}\sum_{i=1}^{n}\nabla\mathcal{Q}_i(\mathbf{w})$, where $\alpha$ is the learning rate used to control the step size of gradient descending. Since the loss function is used to estimate the difference between the predicted value and the ground truth, this equation can be divided into two phases: forward propagation that aims to calculate the predicted value of sample $\mathbf{x}_i$, and backward propagation that calculates the update of $\mathbf{w}$. Calculating gradient over the whole dataset can achieve fast convergence with a small number of iterations, but results in a high computation cost in each iteration. In contrast, selecting a single sample in each iteration is fast for the gradient computation but leads to a slow convergence rate. To address this issue, a small subset of samples (called a batch) are randomly chosen in each iteration, and we can tune the batch size to balance the convergence speed against the overhead.

\textbf{Neural Networks.} Neural networks, or say deep learning, have been widely used in many fields such as computer vision, natural language processing and bioinformatics, thanks to its capability of modeling complex relationships among high-dimensional data. A neural network consists of multiple layers of neurons. Each neuron receives inputs from the previous layer, conducts linear transformation and/or nonlinear mapping, and feeds outputs to the next layer. Neural networks can fulfill multi-class classification tasks with multiple output neurons which constitute the output layer $z_i$, where  \emph{softmax} function $f(z_i)=\frac{e^{z_i}}{\sum_{i=1}^{l}e^{z_i}}$ is leveraged to transform the inputs to a probability distribution among all output categories ($l$ is the total number of categories). The forward propagation of neural networks is similar to that of regression algorithms, and the backward propagation needs to be separated in a chain rule, from the output layer to the input layer step by step.

\subsection{Collaborative Learning}
To avoid privacy leakage of sensitive data distributed among multiple clients, collaborative learning is a new computing architecture that learns a global prediction model without sharing the training data~\cite{chilimbi2014project, dean2012large, lin2017deep, zinkevich2010parallelized}. During the training process, a central server maintains a global model and randomly chooses a subset of clients to update the model in each round. The chosen clients download the current global model and compute an updated model based on their local data, and then return the updates to the server. After receiving results from all the chosen clients, the server aggregates them  (typically by averaging) to construct a new global model. There are two methods to achieve collaborative learning, i.e., uploading the weights directly or the difference between the updated and original model. To keep the consistency of the attacks, we consider collaborative learning achieved by averaging updates rather than weights. Algorithm~\ref{alg:fedavg} provides the pseudocode.

\subsection{Poisoning Attacks}
Generally, poisoning attacks assume the existence of an attacker (e.g., a malicious client) who aims at changing the behaviors of a trained model on specific inputs through manipulating the training dataset, e.g., changing the labels of records, or injecting poisoned training data samples. Concretely, two prevalent types of attacks have been proposed. In \emph{label-flipping attacks}~\cite{bagdasaryan2018backdoor, cao2018vggface2}, the attacker flips the labels of all the original data samples of one class to another class. In \emph{backdoor attacks}~\cite{gu2017badnets, bhagoji2019analyzing}, the labels of data samples with certain features are relabeled, and the features can be exploited to fool the model. The former is preferred for scenarios that data samples with the same labels are similar, e.g., face recognition, and the latter is more suitable for scenarios that samples in a class are diverse, e.g., image tagging.

In collaborative learning, prior works show that the model is vulnerable to poisoning attacks as well~\cite{shen2016uror, fung2018mitigating, bagdasaryan2018backdoor}. The attacker can generate a poisoned model by carefully crafting its local training data or updated model. But the difficulty for poisoning collaborative learning is that averaging the updates performed on the server side may mitigate the effect of the poisoned update on the global model. To achieve a successful attack, two methods have been proposed~\cite{bagdasaryan2018backdoor}. One is scaling the poisoned update by a large factor. If the attacker knows the number of clients and the global learning rate, it can replace the average with the poisoned update as follows:
\begin{equation}\label{eq:backdoor}
\begin{aligned}
 \tilde{L_m^{t+1}} &= \frac{n}{\eta}X-(\frac{n}{\eta}-1)G^t-\sum_{i=1}^{m-1}(L_i^{t+1}-G^t) \\
  &\approx \frac{n}{\eta}(X-G^t)+G^t.
\end{aligned}
\end{equation}
The other is controlling multiple clients through generating sybils to increase the influence of the poisoned updates on the averaged model.

\begin{algorithm}[t!]
\caption{Collaborative learning with update averaging}\label{alg:fedavg}
%{\footnotesize
\begin{algorithmic}[1]
\STATE \textbf{ServerExecutes}:
\STATE Initialize the global model ${w_0}$
\FOR {round $t = 1$ to $T$}
\STATE ${C_t}\leftarrow$ {the server randomly chooses $K$ clients}
\FOR {each client $j$ in $C_t$}
\STATE $\Delta w_{t+1}^k\leftarrow$ \textbf{ClientUpdates}($kj, w_t$)
\ENDFOR
\STATE $w_{t+1}\leftarrow w_t+\frac{1}{K}\sum_{k=1}^K\Delta w_{t+1}^k$
\ENDFOR
\STATE
\STATE \textbf{ClientUpdates}($k, w_0$):
\STATE Download the current global model ${w_0}$
\FOR {iteration $i = 0$ to $I$}
\STATE $b_i\leftarrow$ randomly choose a batch from its local data
\STATE $w_{i+1}\leftarrow w_i -\eta\nabla l(w_i, b_i)$
\ENDFOR
\STATE $\Delta w_{t+1}=w_I-w_0$
\STATE send $\Delta w_{t+1}$ to the server
\end{algorithmic}
%}
\end{algorithm}

\begin{figure*}[t!]
  \centering
  \includegraphics[width=0.95\textwidth]{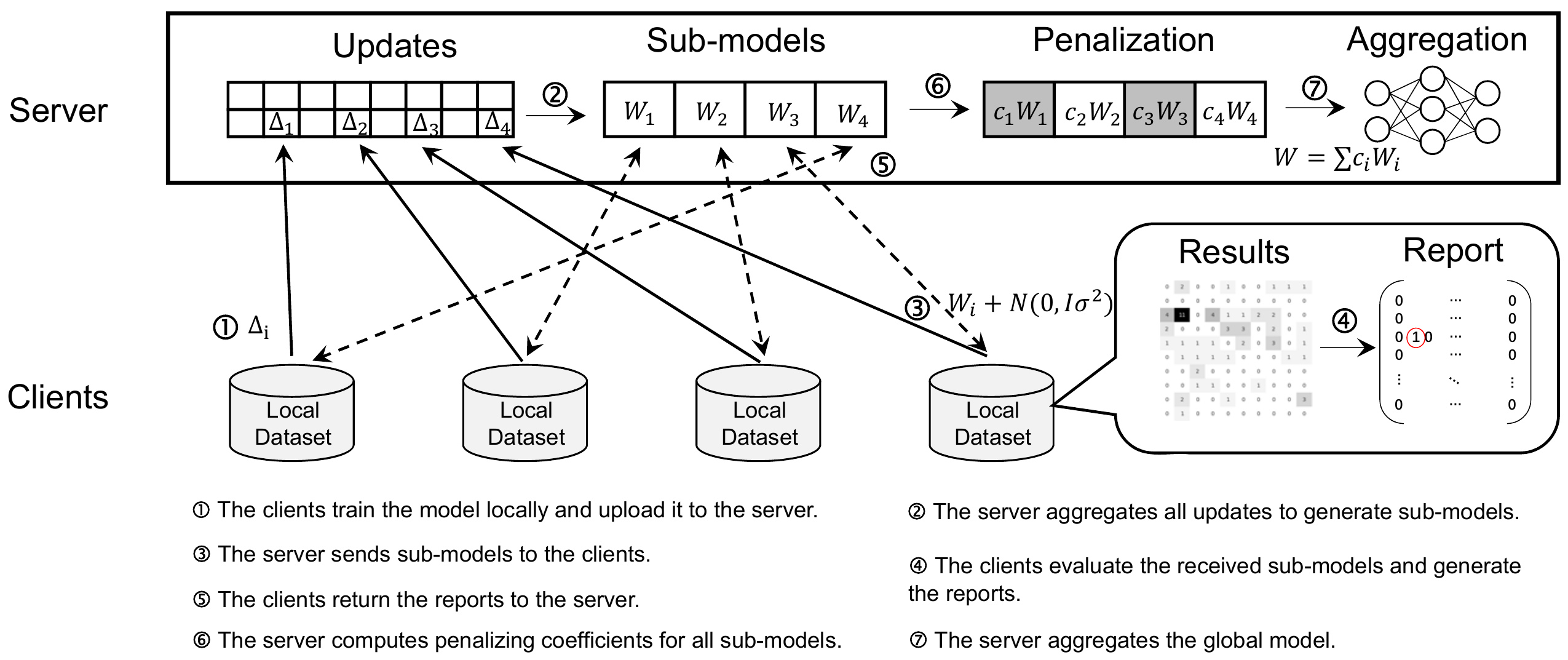}
  %\vspace{-2mm}
  \caption{An overview of our scheme}\label{fig:overview}
  %\vspace{-4mm}
\end{figure*}
\section{Problem Statement and Threat Model}
We consider the generic setting for collaborative learning, where a central server and $N$ clients jointly train a learning model with a given training algorithm. All the training data are kept by clients locally, while the global model is shared among clients for classification or prediction. Our goal is to defend against poisoning attacks (i.e., detecting anomalous updates) carried out by malicious clients. Besides, following prior works~\cite{geyer2017differentially, mcmahan2017learning}, we also aim to prevent malicious clients from inferring the participation of other clients.

We assume that an attacker can fully control a proportion $p$ of clients (sybils) ($0<p<1$), i.e., the attacker can manipulate the training processes by changing the target training algorithm, the training data of the controlled clients, the hyperparameters when updating a model, or directly modifying the parameters of the updates. Following~\cite{bagdasaryan2018backdoor}, we call the primary training task as the \emph{main task}, and the goal of controlling the behavior of the model as the \emph{attacker-chosen subtask} (and we use ``subtask'' for simplicity in the rest of the paper).
%The goal of the attacker is to change the behavior of the model on classifying specific inputs.
We consider a more aggressive (or say concealed) attacker who also tries to aid the main task (i.e., making the global model achieve a high accuracy), instead of  harming it considered in~\cite{blanchard2017machine}. Moreover, it may try to infer an honest client's participation by the differential attack~\cite{geyer2017differentially}.

We assume that the central server is trusted, the outputs of any algorithm running on the server are correct. Besides, the rest $(1-p)N$ clients are assumed to be honest as well, and their model updates  are correctly computed over the honest data. Note that the inputs of benign clients cannot be tampered by the attacker. We consider a varying proportion of malicious clients, and we will show how this proportion affects the detection performance in Section~\ref{sec:results}. More importantly, differing from existing solutions of detecting poisoning attacks~\cite{shen2016uror, blanchard2017machine}, we do not restrict the number of malicious clients and the distribution of the training data. Data can be IID or non-IID across clients.
We only assume that there is a sufficient number of training data such that each class of data is held by multiple clients.

We summarize the important notations in Table~\ref{tab:notation}.

%Moreover, in this paper, we mainly focus on how to detect anomalies in collaborative learning. However, the detection that needs to observe the parameters of uploaded models may leak information about honest client's data. We will discuss how to mitigate this leakage in Section~\ref{}.

\textbf{Types of poisoning attacks.} In this paper, we consider two practical poisoning attacks: label-flipping attacks and semantic backdoor attacks~\cite{bagdasaryan2018backdoor}. Both of these two approaches can be implemented by changing labels without modifying data samples such as the pixels of images. The attack will succeed when the inputs of the victim lie in the poisoned set.
Note that there exists another attack type called pixel-pattern attack~\cite{gu2017badnets, bhagoji2019analyzing}. To introduce a backdoor into the model, however, the attacker needs to modify both the training-time data and the test-time inputs. We emphasize that such an attack is difficult to be deployed in collaborative learning since the test-time inputs belonging to an honest client cannot be compromised.
Therefore we do not consider pixel-pattern attacks in this work.

\begin{table}[!t]
\vspace{1mm}
\normalsize
\newcommand{\tabincell}[2]{\begin{tabular}{@{}#1@{}}#2\end{tabular}}
\centering
\begin{tabular}{c|l}
\toprule[0.8pt]
$\textbf{Notation}$ &  $\textbf{Description}$ \\
\midrule[0.8pt]
$t$ & the $t$-th training round \\
%\hline
$K$ & the number of chosen clients in the current round\\
%\hline
$S_t$ & the set of $K$ clients \\
%\hline%
$L_t^i$ & the update of the $i$-th client \\
%\hline
$w_t$ & the global model in the $t$-th round \\
%\hline
$w_t^i$ & the $i$-th sub-model in the $t$-th round \\
%\hline
$d$ & the number of sub-models in the current round \\
%\hline
$u$ & \tabincell{l}{the number of updates  for generating a sub-model} \\
%\hline
$m$ & the number of sub-models that a client can evaluate \\
%\hline
$e$ & \tabincell{l}{the number of clients that evaluate one sub-model} \\
%\hline
$c_t^i$ & the penalizing coefficient of $w_t^i$ \\
\bottomrule[0.8pt]
\end{tabular}
\caption{Important notations}
\label{tab:notation}
\end{table}

\section{Defending Against Poisoning Attacks: Our Approach}
Intuitively, we observe that if the server has a large test dataset with ground-truth labels, it can easily detect anomalies by feeding honest inputs to the updated model and comparing the results with the ground truth. In collaborative learning, however, the server has no access to the datasets held by clients. Hence our key insight is delegating such a detection process to clients who are able to perform this task. Simply speaking, the server sends the models to a set of selected clients, each of which evaluates the accuracy of the model according to a predefined rule over its local dataset. The server finally adjusts the weights of updates when averaging according to the evaluation results received from clients. Figure~\ref{fig:overview} presents an overview of our scheme.

In this section, we first present our scheme of detecting poisoning attacks in the IID setting, then extend it to the non-IID setting. Finally, we briefly discuss the feasibility and effectiveness of our schemes.

\subsection{Poisoning Attacks in the IID Setting}
Now we present our defense method against poisoning attacks in the IID setting. We propose delegating the detection process to the clients, who can detect abnormal updates through evaluating the performance of the updates over their own data. In an ideal situation, each client measures the performance of all the updated models, which results in the highest success rate for detecting attacks.
In practice, however, downloading and evaluating all the updates are obviously impossible due to unbearable computation and communication costs. In light of this, we propose randomly selecting a small set of clients to evaluate each model. Due to the IID distribution of the training data across different clients, we can still achieve a high detection rate only if the selected clients have sufficient data samples.
%\textbf{Communication efficiency. } The size of the model might be large, therefore downloading multiple models for each client may cause expensive communication costs. Fortunately, we observed that the final layer is the most important role of poisoning, because its parameters will directly affect the output. Consider a simple example that a digit classifier uses a fully-connected layer with the softmax function as the final layer. Only modifying the parameters of the target class can help to output it with a high probability. Since the aim of attacks in federated learning is making the aggregated model be close to the attacked model, here we replace the final layer of a normal model with an attacked one to confirm this observation.

%Figure~\ref{} shows the accuracies of both the label-flipping attack and the backdoor attack on multiple tasks.

\begin{algorithm}[t!]
\caption{IID Delegation}\label{alg:allocate_iid}
%{\footnotesize
\begin{algorithmic}[1]
\STATE Randomly shuffle the $K$ updates
\STATE $W_d\leftarrow$ Average every $u$ updates to obtain $d$ sub-models
\FOR {each update $i\in W_d$}
\FOR {$j=1$ to $e$}
\STATE Randomly choose a client $k$ from $S_t$
\IF {$w_t^i$ is not generated by $k$'s update and the number of sub-models assigned to $k$ is less than $m$}
\STATE Assign $w_t^i$ to $k$
\ELSE
\STATE Randomly choose another client from $S_t$
\ENDIF
\ENDFOR
\ENDFOR
\end{algorithmic}
\end{algorithm}

\textbf{Delegating detection task.}
From the perspective of security and efficiency, the strategy of allocation needs to satisfy the following three requirements: (1) any update should not be assigned to its owner. Otherwise, it is possible for an attacker to evade the detection by forging evaluation results; (2) the number of  models assigned to the same client should not exceed a predefined threshold. If a client needs to evaluate much more uploads than others, waiting for its results may cause a significant delay in this round; (3) the communication costs should be as small as possible.

The first two requirements can be easily satisfied by adding the corresponding constraints in the allocation mechanism. For the third requirement, we propose aggregating a batch of updates into a sub-model for evaluation rather than dealing with them individually. Specifically, the server randomly partitions $K$ updates into $d$ parts. Then each part will be aggregated into a sub-model by averaging $u=\frac{K}{d}$ updates. Every sub-model will be randomly assigned to $e$ clients, and every client will evaluate $m$ ($m<d$) sub-models. We mainly consider two types of attacks. One is that some samples are misclassified into a certain class with high probability, and the other is that the accuracy of the model is degraded significantly. Both of these two attacks can be detected by evaluating the performance of the sub-models on certain classification tasks.

Algorithm~\ref{alg:allocate_iid} provides the pseudocode of delegating the detection task to the clients.

\textbf{Model aggregation.} In general, the model updates are aggregated by averaging. The server can adjust the weights in the averaging based on the clients' evaluation results, such that an update with a higher probability of being poisoned has a lower impact on the global model. For the weight adjustment, a naive approach is letting the client report the number of correctly classified data samples when evaluating the update. However, this method is vulnerable to a malicious client who reports an arbitrarily large value for a poisoned model in order to add its weight. Therefore, we propose using the strategy of majority voting, where the clients are only required to submit a binary matrix claiming whether the corresponding category of data samples is correctly classified.

According to the binary matrix, the server is easy to summarize the count that an update has been reported to be potentially malicious. To adjust the weights of the updates, we add a penalizing coefficient for each update based on the counts. We observe that such a coefficient should be as small as possible when the count is low due to the potential false-positive reports of the updates. We design a hierarchical method to determine the coefficient $c$ according to the returned results as follows. We first give three observations: (1) with the increasing of anomaly reports, the coefficient should reduce sharply when the reports have higher confidence, (2) if more than half of the clients report the anomaly for the same sub-model, it should be discarded in the aggregation and (3) if there is only one client reports the anomaly, the penalizing coefficient should not be too large as the result might be inaccurate. Based on these observations, we propose the following penalizing function

\begin{equation}
\label{eq:coefficient}
  c_t^i=\max(\frac{-4v}{(e-2)^{2}}(r_t^i)^2+\frac{8v}{(e-2)^{2}}r_t^i-\frac{4v}{(e-2)^2}+v, 0),
\end{equation}
where $r_t^i$ is the number of clients who report the $i$-th update to be malicious in round $t$, and $v$ is the initial penalizing coefficient when only one anomalous update is reported (i.e., $r_t^i=1$).
In Section~\ref{sec:results}, we will show that setting $v$ to be 0.5 is effective in mitigating poisoning attacks.

In summary, the model aggregation in our scheme is formulated as
 \begin{equation}
   w_{t+1}=\frac{1}{d}\sum_{i=0}^dc_{t+1}^iw_{t+1}^i.
 \end{equation}
The pseudocode of this process is shown in Algorithm~\ref{alg:detection}.

\begin{algorithm}[t!]
\caption{Detection in IID setting}\label{alg:detection}
%{\footnotesize
\begin{algorithmic}[1]
\STATE \textbf{ServerExecutes:}
\STATE Initialize the global model ${w_0}$
\FOR {round $t = 1$ to $T$}
\STATE ${S_t}\leftarrow$ Randomly choose $K$ clients
\FOR {each client $k$ in $S_t$}
\STATE $L_{t+1}^k\leftarrow$ \textbf{ClientUpdate}($k$,$w_t$)
\ENDFOR
\STATE \textbf{IID Delegation}
\FOR {every client who performs the detection}
\STATE $r_k\leftarrow$ Return the binary matrix of the evaluation results
\ENDFOR
\FOR {$i=1$ to $d$}
\STATE Compute the penalized coefficient $c_{t+1}^i$
\ENDFOR
\STATE $w_{t+1}\leftarrow \frac{1}{d}\sum_{i=0}^dc_{t+1}^iw_{t+1}^i$
\ENDFOR

%\EndFunction
\end{algorithmic}
%}
\end{algorithm}

\subsection{Poisoning Attacks in the non-IID Setting}
To extend our detection scheme to the non-IID setting, we only need to  modify our client selection strategy, and the model aggregation design remains unchanged since it is completed on the server side which is irrelevant to the distributions of the data.

\begin{figure}[t!]
  \centering
  \includegraphics[width=0.95\columnwidth]{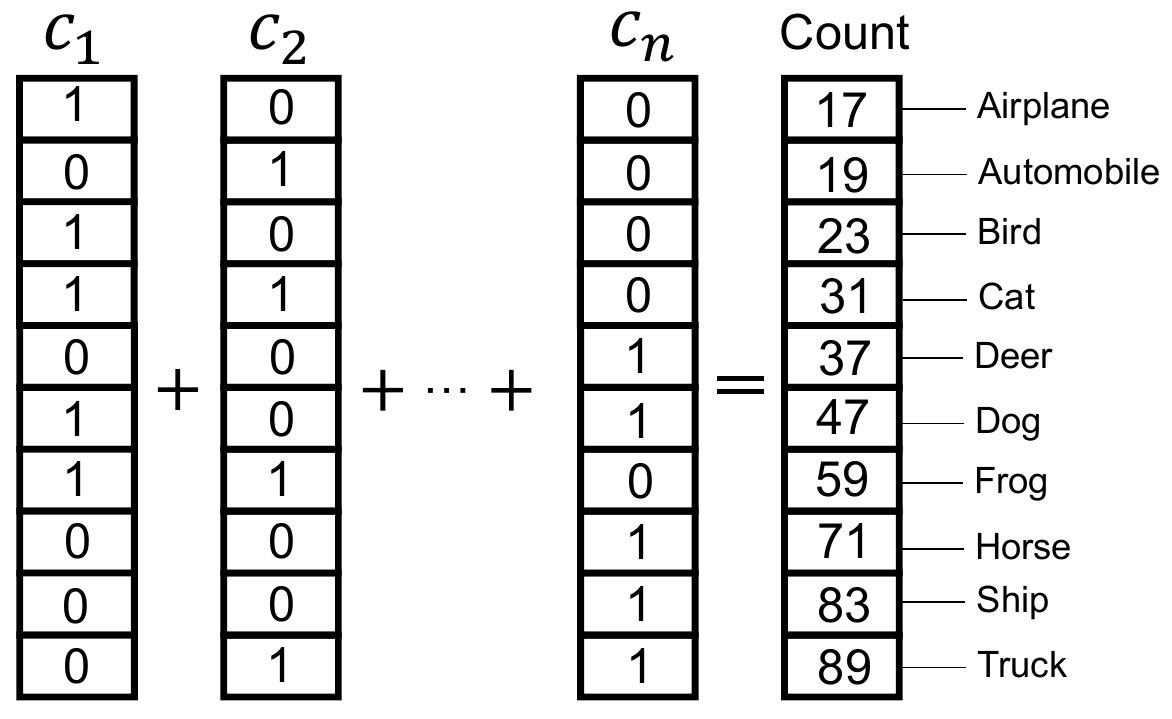}
  %\vspace{-2mm}
  \caption{Collecting information about the distributions of clients' data}\label{fig:count}
  %\vspace{-4mm}
\end{figure}

\textbf{Delegating detection task.} The main difference between the IID and non-IID settings is the imbalance between classes in clients' datasets. Some classes of data samples may only exist in a part of clients. Therefore, the models trained by these clients cannot be directly evaluated over other clients' data. To tackle this problem, we design a protocol to dynamically distribute sub-models to suitable clients for evaluation, where the clients will only evaluate the sub-models over their data classes.
%Since the server has no access to clients' data, it is necessary to rely on the information clients provided to determine the strategy of allocating clients.
Specifically, we let each client first tell the server if the number of data samples for each label is larger than a threshold, such that the server can assign models to the clients who have sufficient samples of the same class. At the beginning of each round, this information will be updated by the clients due to the possible changes in the local data. It can be represented with a binary vector where 1 means that the client has sufficient data for a given class, e.g., it has more than 100 car images, otherwise 0. Figure~\ref{fig:count} presents an example of this process. In each round, based on the updated information about the local data, the server will select a set of new clients who have the datasets with the most similar classes.

Another problem for the non-IID setting is that there lacks the sufficient number of clients who can evaluate samples of some classes. As a result, it is hard to balance the workloads and some clients will be assigned too many evaluation tasks, causing a significant delay. More specifically, in each round, if $d>em$, the number of evaluators is not enough to detect all the sub-models. To solve this problem, before delegating tasks, the server first sums the binary vectors received from clients and sorts the result in descending order, and then sets $d$ to be the smallest value in the resulting vector. Finally, $K$ updates are aggregated into $d$ sub-models.

%To solve this problem,
%%we aggregates all uploads to $m$ parts to match these clients.
%in each round, the server firstly sums all the vectors of clients' data up to obtain the statistics about each class of data is held by how many clients, then sorts the numbers from smallest to largest and sets the minimum as $m$ and aggregates the $K$ uploads to $d$ parts.

\begin{figure}[t!]
  \centering
  \includegraphics[width=0.95\columnwidth]{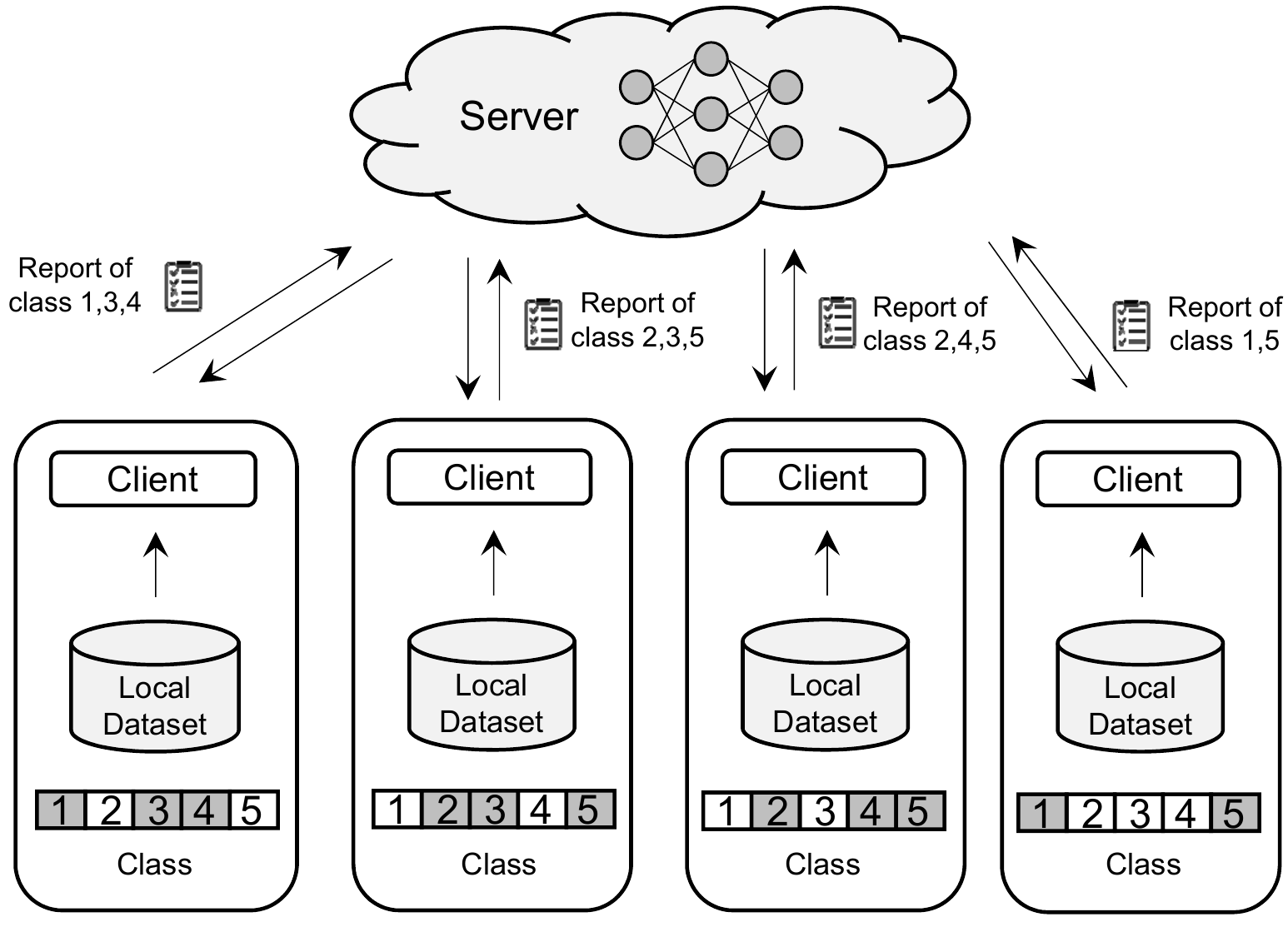}
  %\vspace{-2mm}
  \caption{An example of allocating detection tasks in the IID setting}\label{fig:allocation}
  %\vspace{-4mm}
\end{figure}

Note that if some clients have multiple classes of data, its detection results can be reused to reduce the communication cost of assigning a new candidate. Figure~\ref{fig:allocation} depicts a brief example for this process. For a sub-model $w_i$, the server assigns all the evaluation tasks to 4 clients while every class is evaluated by two clients (i.e., $e=2$). Since the first client holds the data for digit 1, 3 and 4, it is assigned to evaluate the accuracy of the sub-model that classifies these three digits. Then the server only needs to choose another one who holds these classes of data, rather than choosing two new clients. For example, for digit 1, we will assign the evaluation to the first and the fourth clients. Algorithm~\ref{alg:allocate_noniid} provides the pseudocode of delegating detection tasks in the non-IID setting.
%Note that if the more labels the dataset has, or the higher similarities of data with the same label, the better performance of our scheme will have as the allocation will be more accurate.

\subsection{Discussions. }
In this section, we discuss the feasibility and effectiveness of our defense method.

\textbf{Feasibility.} Our design is based on the assumption that there exist honest clients with sufficient data samples to evaluate the updates. Since one of the motivations for collaborative learning is utilizing more data collected from diverse clients to improve the performance of the model, our assumption is easy to be satisfied. In this paper, we concern about two representative scenarios of collaborative learning. The first one is the cooperation of multiple institutions, e.g., training a diagnostic model between multiple hospitals, or constructing a recommendation system between multiple malls. In this case, allocating every sub-model to several clients is enough to ensure the correctness of the detection, since each client is able to collect diverse data. The other scenario is federated learning that aggregates massive individual users' data~\cite{mcmahan2017communication}, e.g., training a model for mobile keyboard prediction~\cite{konevcny2016federated}. In this case, a very large number of clients are involved since each client does not have enough data set to train an accurate model. These two application scenarios further demonstrate the feasibility of our assumption.

%Even in this case, the communication cost of our scheme is still practicable thanks to the strategy of evaluating sub-models. We will give more details about the communication cost in Section~\ref{sec:results}.

%By contrast, the plausible pixel-pattern backdoors~\cite{gu2017badnets},~\cite{bhagoji2019analyzing} which require the attacker to modify the raw test data may not be detected by our scheme, since it is hard for benign clients to trigger the poisoning target. However, due to the difficulty of controlling the behaviors of benign users by the attacker, this requirement makes such a test-time attack unrealizable in collaborative learning. Note that adversarial examples~\cite{goodfellow2014explaining},~\cite{papernot2017practical},~\cite{kurakin2016adversarial} have a similar effect with the pixel-pattern backdoor that making the model misclassify a well-designed input, but adversarial examples aims to attack a given model rather than break the integrity of a model held by the other.

\textbf{Effectiveness.} The effectiveness of the proposed method is derived from the observation that a large scaling factor will decrease the accuracy of the main task, and a small scaling factor will make the attack fail. Hence, evaluating the accuracy of the model might be an appropriate way to detect if there exists an efficient attack. While only aggregating several updates into a sub-model, the original scaling factor will be no longer valid unless it precisely mathces the sub-model's size. However, this will significantly affect the success rate of the attack.

The detection performance is related to the number of malicious clients. If there is only a single malicious client, the poisoning attack can be easily detected since most of clients, who are honest participants, will report the bad performance of the poisoned update. While the attacker controls multiple clients by creating sybils, the detection performance might be degraded since the poisoned updates may evade detection when they are assigned to another malicious client who reports false results.
Let $p$ denote the proportion of malicious clients. In each round, the server randomly samples $K$ different updates and aggregates $u$ updates to obtain a sub-models. Each sub-model is delegated to $e$ clients. The probability $p_{evd}$ that a poisoned sub-model is delegated to $t$ malicious clients is

\begin{equation}\label{equ:probability}
  p_{evd} = \sum_{i=1}^u\frac{\binom{K(1-p)}{u-i}\binom{Kp}{i}}{\binom{K}{u}}\times\frac{\binom{Kp-i}{t}\binom{K(1-p)-u+i}{e-t}}{\binom{K-u}{e}}.
\end{equation}
For example, if the server chooses 100 clients in this round with 10 malicious clients included, and each update is delegated to 3 clients. The probability that a sub-model which includes one poisoned update is delegated to one malicious client is 0.12. The probability that it can evade the detection, i.e., all the 3 clients who evaluate the update are all malicious is 0.0003, which is negligible. In fact, finding a part of poisoned updates, rather than all of them, would be enough to mitigate the poisoning attack. In summary, our scheme remains effective even if multiple clients are compromised by the attacker. We will show how the number of sybils affects the detection performance in Section~\ref{sec:results}.

We note that our framework can be combined with the existing defense solutions against sybil attacks such as FoolsGold~\cite{fung2018mitigating}, which measures the cosine similarity between updates and discards the ones that are significant similarity. Specifically, the server can run FoolsGold first to identify sybils, and then adjust the allocation strategy to reduce their weights. Note that we can modify the cosine similarity of multiple malicious clients by adding orthogonal noise vectors to their updates. As the noises will decrease the accuracy of the model significantly, the honest clients can detect abnormal updates more easily. Moreover, methods~\cite{wang2019neural, liu2018fine} which aim at detecting poisoning attacks for a trained model can be introduced to enhance the detection performance. But, due to the large overhead of evaluating every update, the proposed strategy of aggregating updates as sub-models is also needed.

\section{Privacy Protection}
As indicated in~\cite{geyer2017differentially, mcmahan2017learning}, directly revealing aggregated updates to the adversary may reveal the information about whether a client joins the learning process, which is called client-level privacy. Moreover, compared with computing global model, the sub-model is derived from fewer clients. In light of this, we propose integrating differential privacy to our scheme to protect client-level privacy.

\begin{algorithm}[t!]
\caption{non-IID Delegation}\label{alg:allocate_noniid}
%{\footnotesize
\begin{algorithmic}[1]
\STATE $v\leftarrow$ Record clients who hold the data of each class
 in $S_t$
\STATE $num\leftarrow$ Count the number of clients for each class
\STATE $m\leftarrow$ Find the minimum number in $num$
\STATE Randomly shuffle all the $K$ updates
\STATE $W_d\leftarrow$ Average every $u$ updates to get $m$ sub-models
\FOR {each sub-model $w_i\in W_d$}
\FOR {each class $j$}
\FOR {$k=1$ to $|v_j|$}
\STATE Randomly choose a clients $k$ from $v_j$
\IF {$w_i$ is not generated by $k$'s update and the number of sub-models assigned to $k$ is less than $e$}
\STATE Assign $w_i$ to $k$
\ELSE
\STATE Randomly choose another client from $v_j$
\ENDIF
\ENDFOR
\ENDFOR
\ENDFOR
\end{algorithmic}
\end{algorithm}

\textbf{Protecting client-level privacy. }The server can disguise the sub-models by injecting Gaussian noises directly, and send the perturbed sub-models to the clients. The clients will evaluate the performance of the perturbed sub-models and return the evaluation results, based on which the server adjusts the weights of the updates. Since all the updates in a round are partitioned to multiple disjoint sets before adding noises, the above process only spends the privacy budget once according to the parallel composition theorem~\cite{dwork2006our, dwork2009differential}. Besides, our experimental results in Section~\ref{sec:results} show that  adding noises to the sub-models has a negligible impact on the accuracy of detecting anomalous updates.

Note that it is necessary to disguise the global model before publishing. Since every update is accessed twice, it will double the privacy cost in each round. Fortunately, the total privacy cost can be bounded in a tight range to obtain a meaningful privacy guarantee by using the moments accountant~\cite{abadi2016deep}. And the noises have a negligible impact on the performance of the global model as well, as shown in~\cite{geyer2017differentially, mcmahan2017learning}.

\begin{figure*}[!th]
\begin{minipage}[t]{0.98\textwidth}
\centering
  \subfigure[MNIST]{
  \includegraphics[width = 0.32\textwidth]{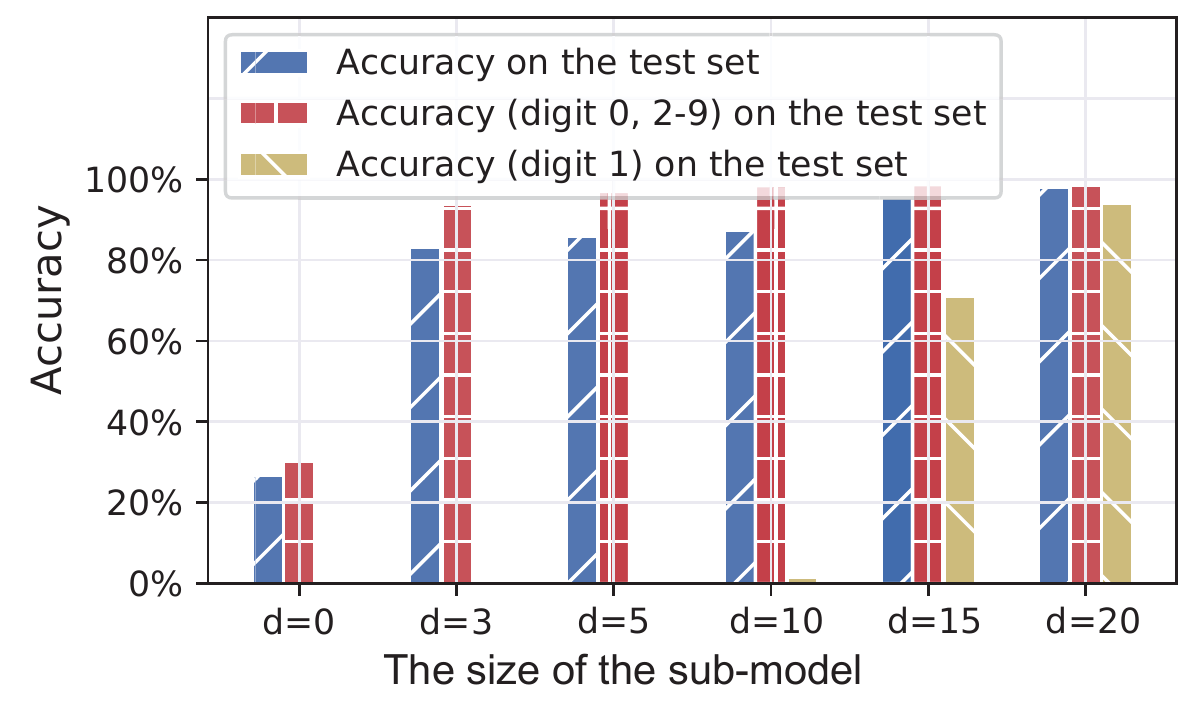}}
  \subfigure[KDDCup]{
  \includegraphics[width = 0.32\textwidth]{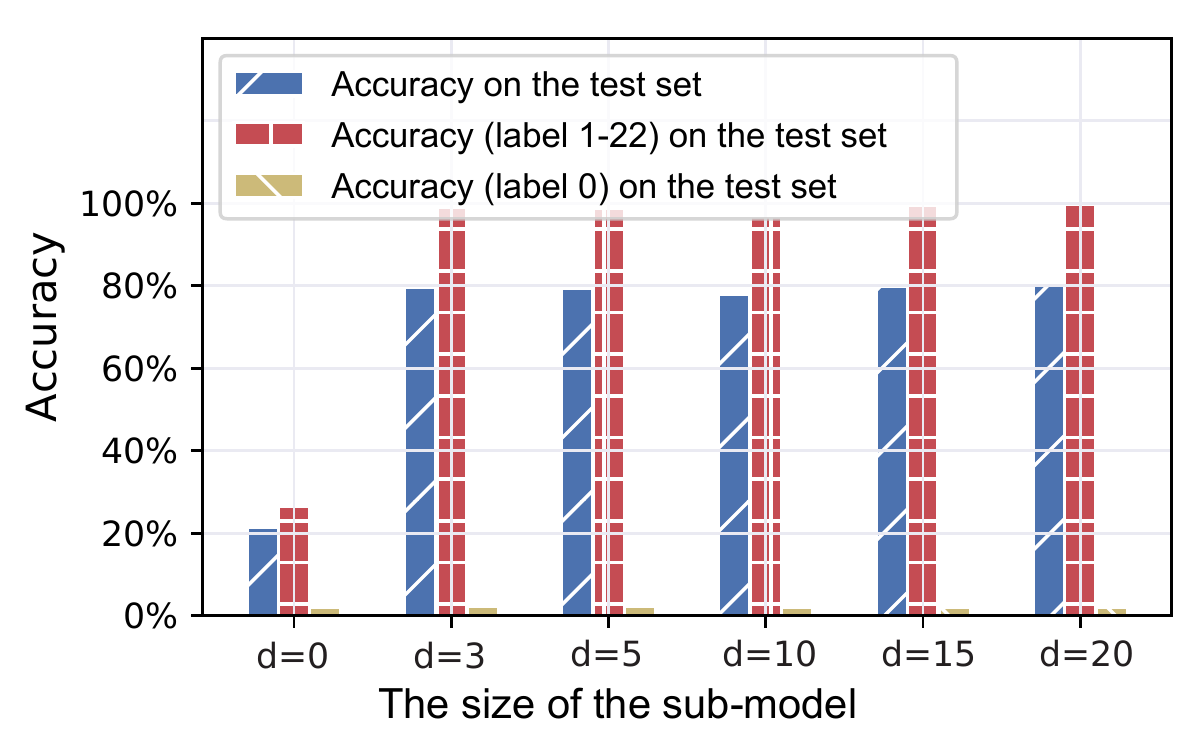}}
  \subfigure[CIFAR-10]{
  \includegraphics[width = 0.32\textwidth]{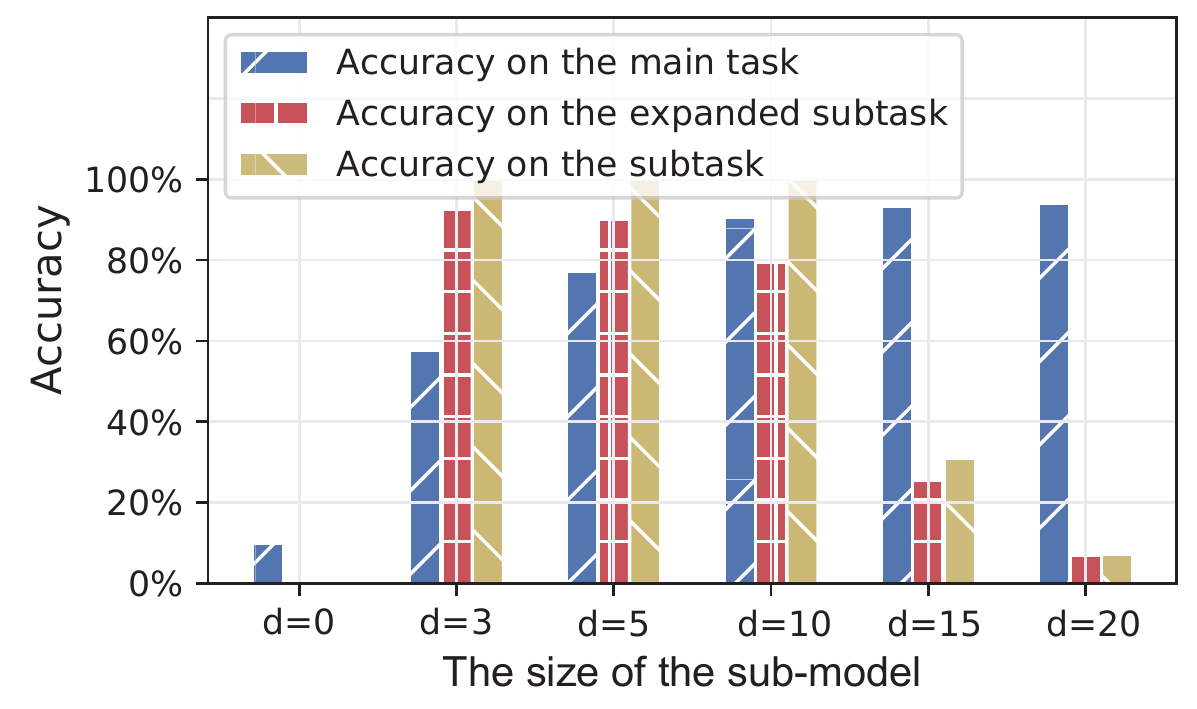}}
  \caption{The effect of the sub-model's size on its accuracy }\label{fig:submodel}
\end{minipage}
%\vspace{-3mm}
\end{figure*}

\begin{figure*}[!th]
\begin{minipage}[t]{0.98\textwidth}
\centering
  \subfigure[MNIST]{
  \includegraphics[width = 0.32\textwidth]{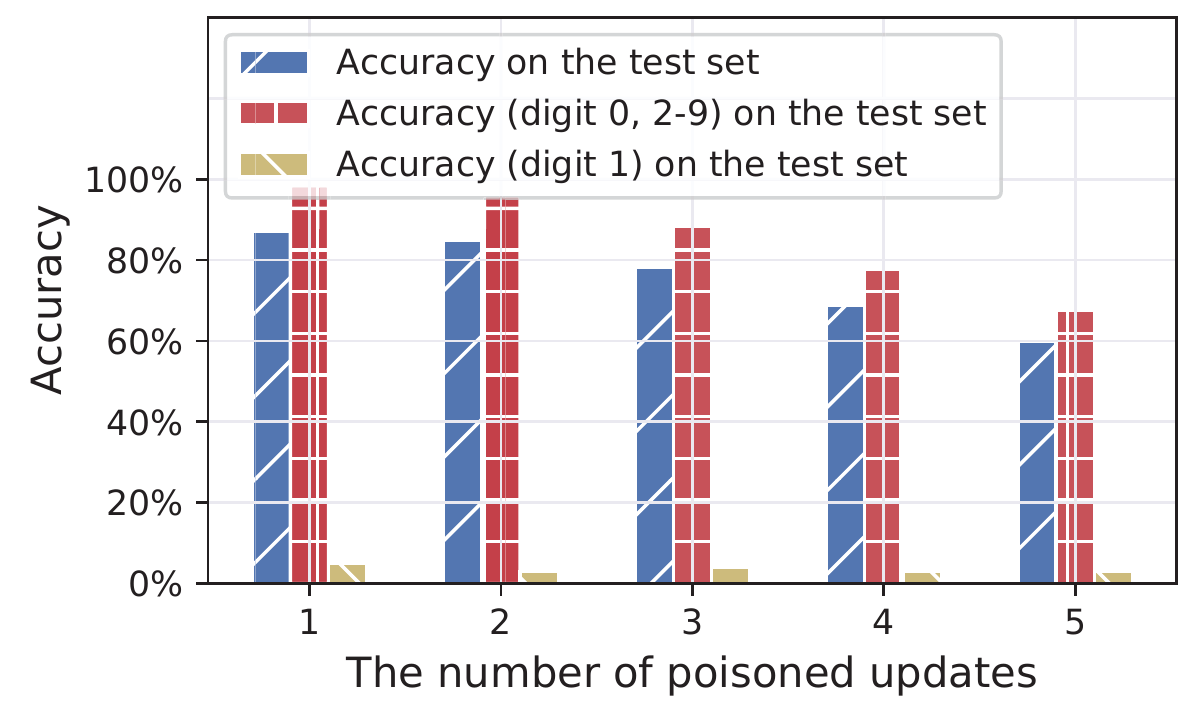}}
  \subfigure[KDDCup]{
  \includegraphics[width = 0.32\textwidth]{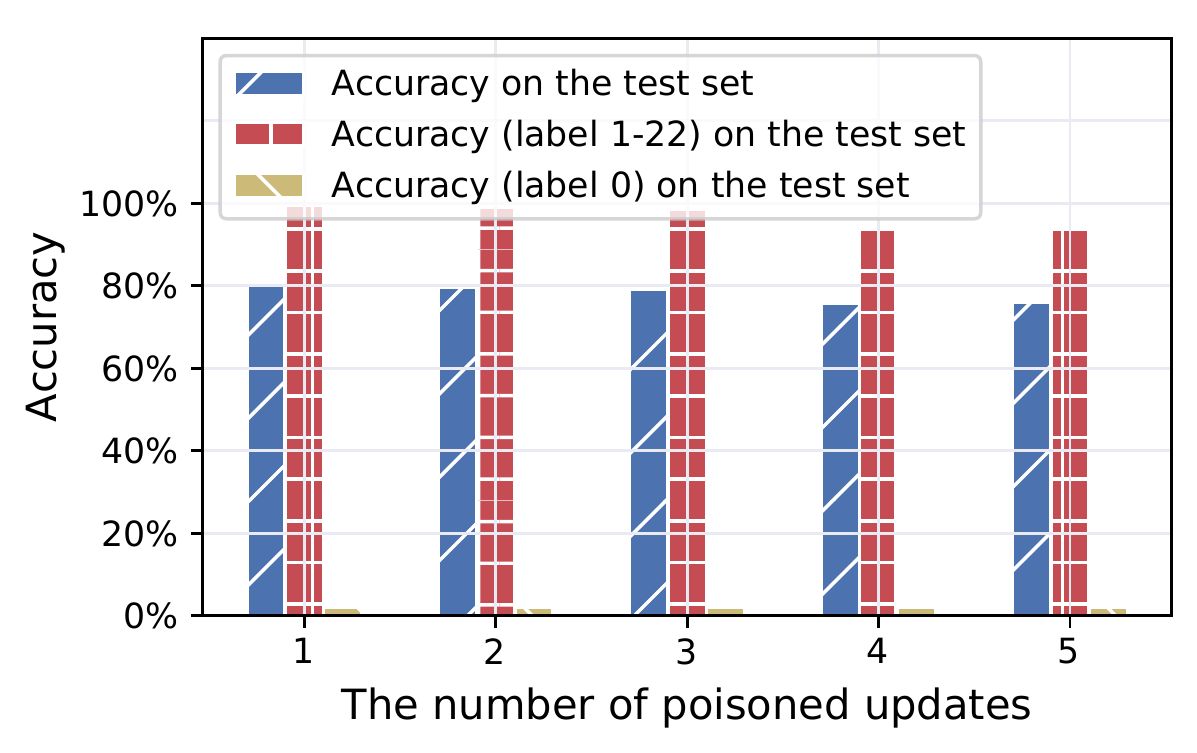}}
  \subfigure[CIFAR-10]{
  \includegraphics[width = 0.32\textwidth]{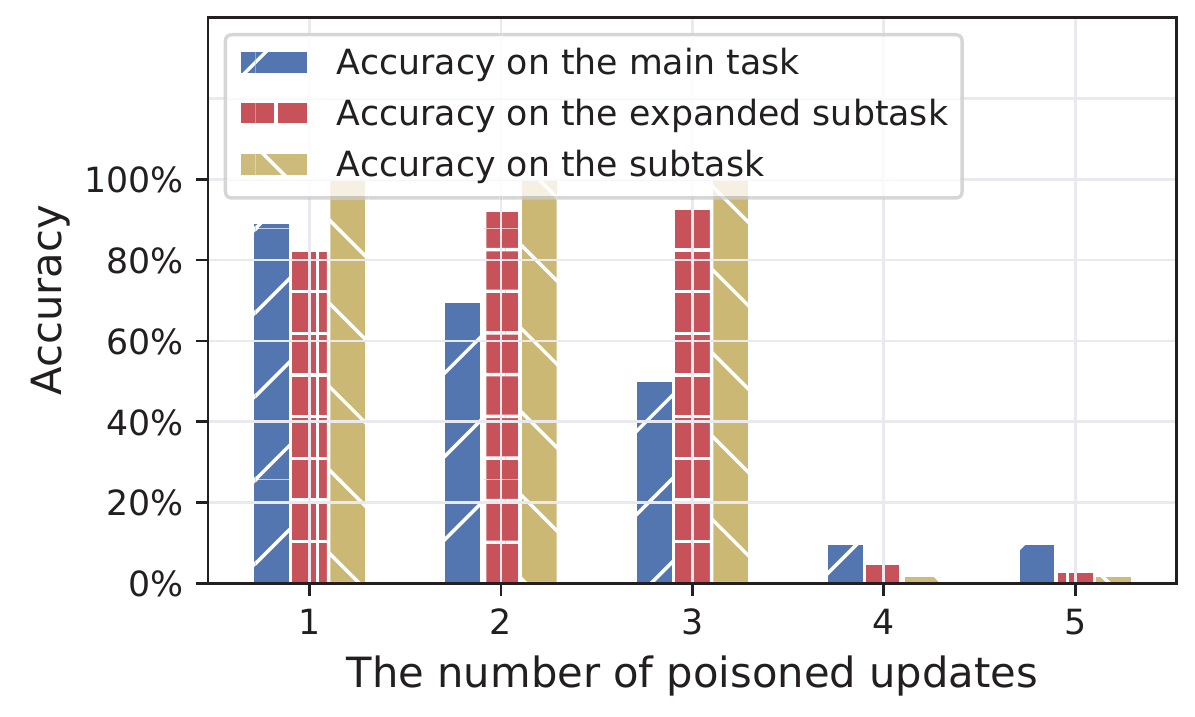}}
  \caption{The effect of the number of poisoned updates on the sub-model's accuracy}\label{fig:mal_submodel}
\end{minipage}
%\vspace{-3mm}
\end{figure*}

\textbf{Privacy vs. Detection Rate. } The number of updates that are aggregated into sub-models is a major factor for the performance of detection.  Intuitively, using fewer updates to generate the sub-model makes it easier to detect which update has been poisoned. However, this enables the attacker to infer the content of the updates from a smaller set, leading to the increase in the risk of privacy leakage. We thus observe that there exists a trade-off between the detection rate and the privacy guarantee.  According to the implementation of client-level privacy~\cite{geyer2017differentially}, we have
\begin{equation}
	w_{t+1}=w_t+\frac{1}{K} (\sum_{i=0}^{K}\Delta w^i/\max (1,\frac{||\Delta w^i||_2}{S})+\mathcal{N}(0,\sigma^2S^2)).
\end{equation}
With the decreasing of $K$, i.e., the number of updates in aggregation, the influence of poisoned update on the sub-model is increasing, making it easier to detect them. In contrast, to keep the usability of the sub-model, the noise should be maintained in a low range, which causes a weaker privacy guarantee.  We will evaluate the impact of differential privacy guarantee on the detection rate in Section~\ref{sec:results}.

\section{Evaluation}
We implement a prototype of our scheme with Tensorflow~\cite{Tensor}.  All the experiments are performed on a workstation with Ubuntu 16.04, Intel Xeon W-2133 CPU, 64 RAM and an NVIDIA 2080Ti GPU card.

\subsection{Datasets}
Since the label-flipping attack and the semantic-backdoor attack are designed for different tasks, we evaluate our scheme for these two attacks on two different datasets. For the label-flipping attack, we use two datasets MNIST~\cite{MNIST} and KDDCup~\cite{KDDCup} used in~\cite{fung2018mitigating}. The former is a well-known handwritten digits dataset which consists of a training set with 60,000 examples, a test set with 10,000 examples. Each example consists of 28x28 pixels and a label. The pixel values are in the range of [0, 255], and all the values are normalized into the scope of [0,1]. The later has 494,020 examples with 23 classes and 41 features. For the semantic-backdoor attack, we use CIFAR-10~\cite{CIFAR10} as the same with~\cite{bagdasaryan2018backdoor}. It consists of 60,000 32x32 color images in 10 classes and every class includes 6,000 examples. 50,000 examples are partitioned into the training set and the others belong to the test set. All the pixel values are also normalized into [0,1].

\subsection{Setup}
We construct a CNN with two convolutional layers with 32 and 64 channels respectively, a fully connected layer with 512 neurons and a softmax output layer for the MNIST task. Both of the two convolutional layers are followed with a 2x2 max pooling layer. In total, it has 1,663,370 parameters. For KDDCup, we train a fully-connected network which has one batch normalization layer, and one fully-connected layer with 23 neurons and the ReLU activation function. The task for CIFAR-10 uses the lightweight ResNet18 model~\cite{he2016deep}, which has about 2.7 million parameters and consists of 17 convolution layers and one fully-connected layer. All the benign clients perform the training algorithm and set the batch size as 64, using the GradientDescentOptimizer with the default learning rate.

For MNIST, the label-flipping attack is implemented by changing all the labels from 1 to 5 on the malicious clients' datasets. For KDDCup, we mis-label 2\% samples with the ``Normal'' class like the prior work~\cite{fung2018mitigating}. In particular, we enhance the prior label-flipping attack by scaling the poisoned update as did in the backdoor attack. The semantic-backdoor attack is implemented in the same way as with~\cite{bagdasaryan2018backdoor}. Examples of cars with racing stripes, vertical stripes on background wall and painted in green are relabeled as birds. Moreover, we expand the dataset with 1,000 randomly rotated and cropped versions of backdoor samples. And we measure the success rate of the attack on the original and the extended datasets respectively.

To simulate the IID setting, we randomly divide all the normal examples into $N$ parts where $N$ is the number of clients. For the non-IID setting, we follow the method proposed in~\cite{mcmahan2017communication} that divides the training set into $2N$ shards where data examples of each shard have the same label, and then assigns each client with only 2 shards. Then we add the poisoned samples into all the datasets of malicious clients. The performance of the model is evaluated by measuring its accuracy in every class.

\subsection{Experimental Results}\label{sec:results}
\textbf{The performance of the sub-model. }In our design, several updates are averaged to obtain a sub-model. To show the effect of sub-models on the detection rate, we assume that each client only involves one poisoned update, and vary the number of honest updates from 1 to 9. The accuracy of the honest updates on the main task is about 93\%.

According to Equation~(\ref{eq:backdoor}) which scales up the poisoned update, we can find that the magnitude of the poisoned update depends on the scaling factor (i.e., the estimated proportion of global learning rate and the number of involved clients). Since the poisoned update is scaled up significantly in general, if the server aggregates the sub-module with fewer updates than the guessed scaling factor, its magnitude will be too large to preserve the accuracy of the global model. Figure~\ref{fig:submodel} presents the accuracy of the main task and the subtask with different sizes of the sub-model. Here we set the scaling factor $n$ to be 10. The attacker can achieve satisfied accuracy on both tasks simultaneously only when the scaling factor is the same as the size of the sub-model. With the increasing of honest updates, the accuracy on the main task will increase as honest updates have larger influences than the poisoned updates. Moreover, when the scaling factor is related to the number of clients in generating the sub-model, aggregating sub-models will further reduce the influence of the poisoned update. If the scaling factor is larger than the number of honest updates, the accuracy of the sub-model decreases significantly, which makes it easier to detect anomalous updates.

From Figure~\ref{fig:mal_submodel} we can see that the accuracy on the main task will decrease when more poisoned updates are chosen to generate the sub-model. This is because more scaled updates will expand the magnitude of the sub-model. If there are too many poisoned updates, the accuracy on the main task and the subtask decreases dramatically. We owe it to the fact that excessive magnitude destroys the usability of the model completely. In summary, if a sub-model includes more poisoned updates, it has a higher probability to be detected.

\begin{figure*}[!th]
\begin{minipage}[t]{0.98\textwidth}
\centering
  \subfigure[Accuracy of the label-flipping attack]{
  \includegraphics[width = 0.32\textwidth]{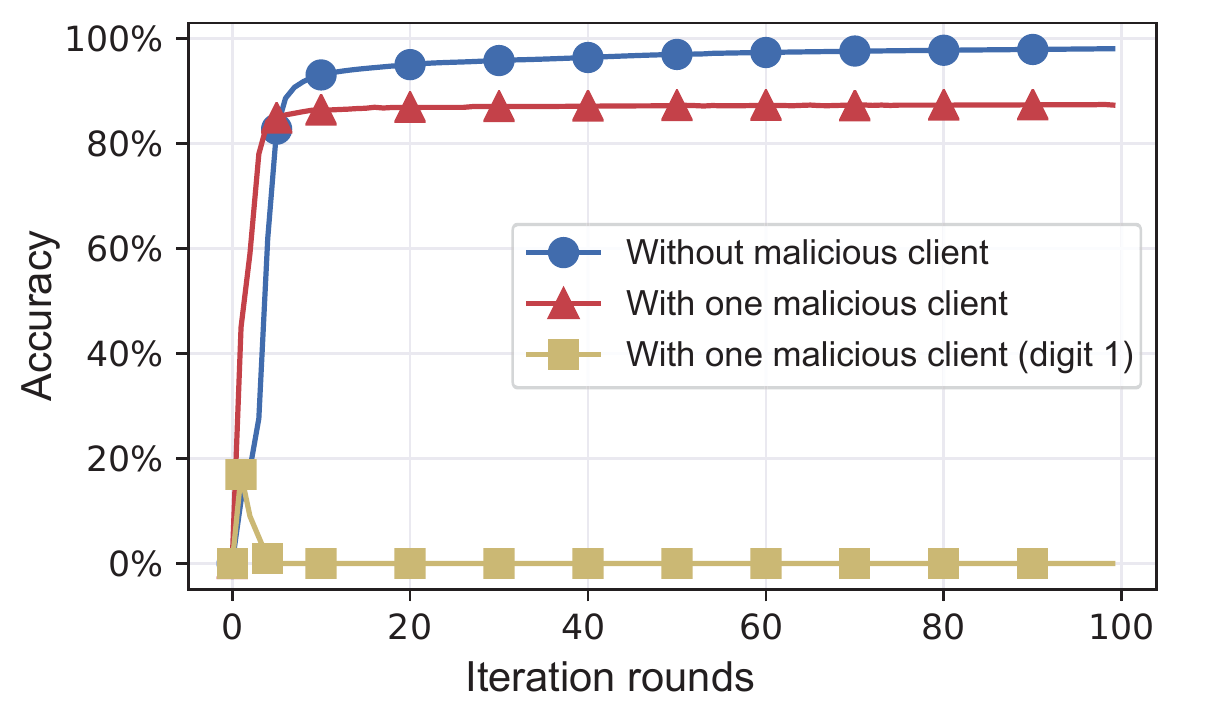}}
  \subfigure[Main task accuracy of the semantic backdoor attack]{
  \includegraphics[width = 0.32\textwidth]{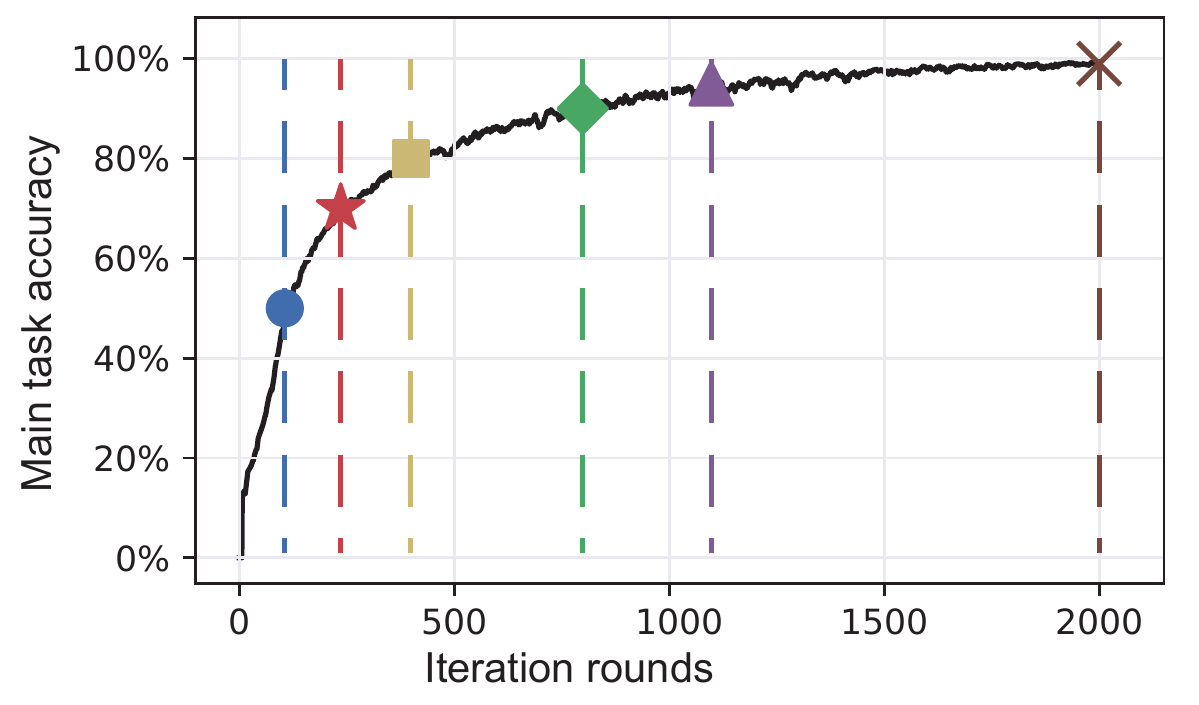}}
  \subfigure[Subtask accuracy of the semantic backdoor attack]{
  \includegraphics[width = 0.32\textwidth]{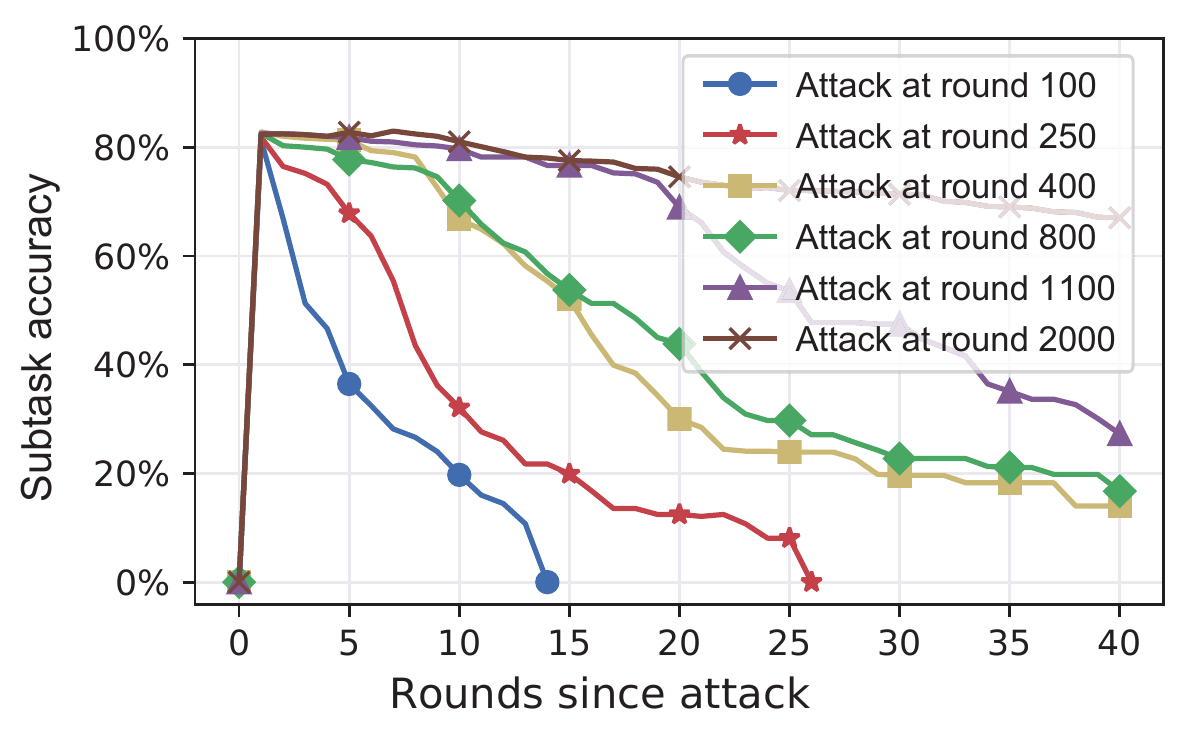}}
  \caption{The accuracy of the subtask when performing the attack}\label{fig:iteration}

\label{fig:iteration}
\end{minipage}
%\vspace{-3mm}
\end{figure*}

\begin{figure*}[!th]
\begin{minipage}[t]{0.98\textwidth}
\centering
  \subfigure[MNIST]{
  \includegraphics[width = 0.32\textwidth]{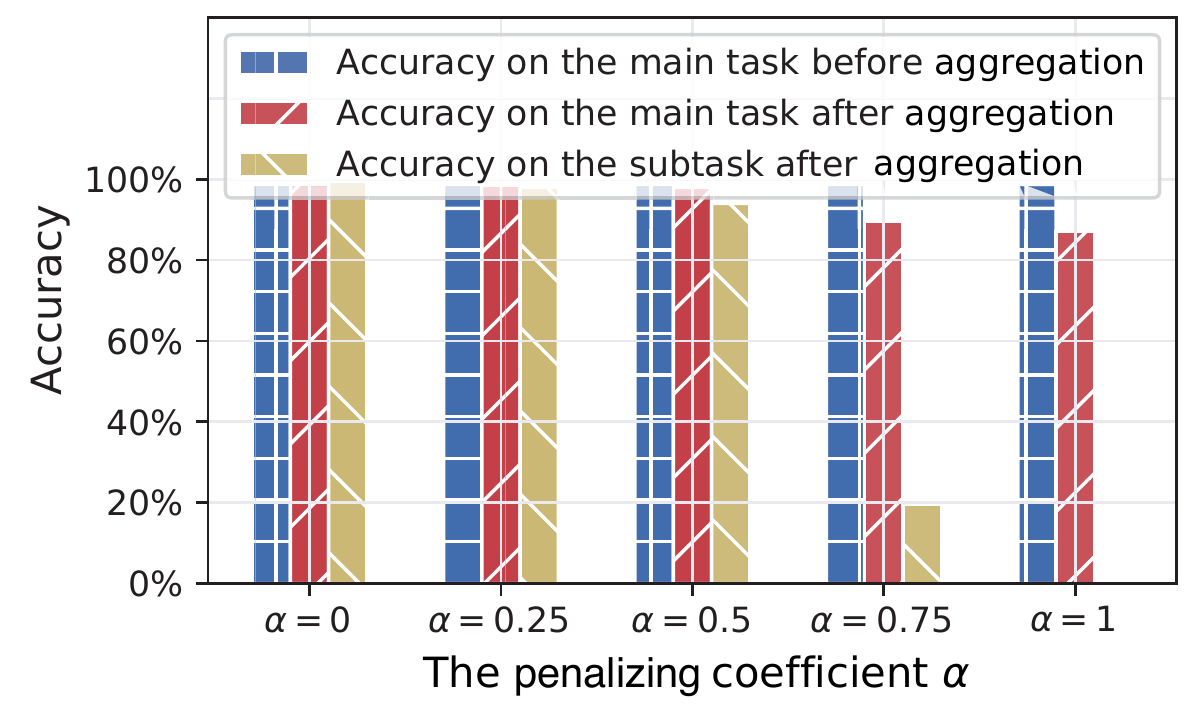}}
  \subfigure[KDDCup]{
  \includegraphics[width = 0.32\textwidth]{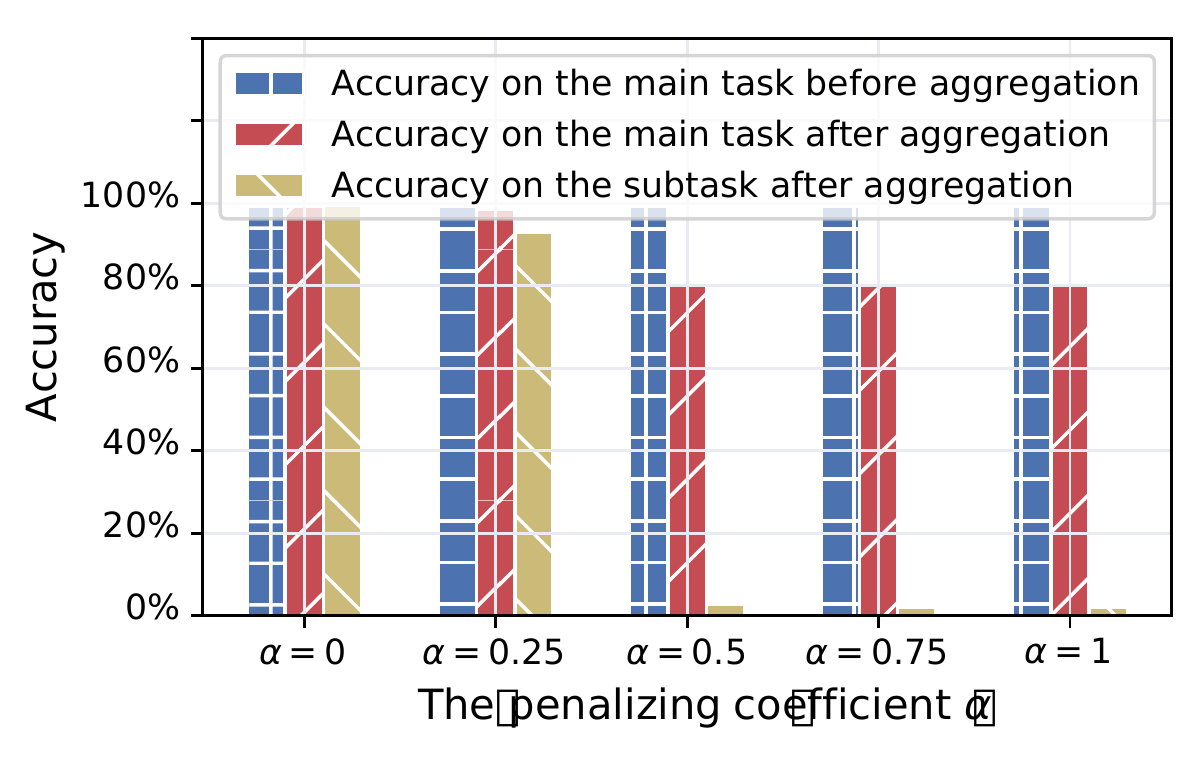}}
  \subfigure[CIFAR-10]{
  \includegraphics[width = 0.32\textwidth]{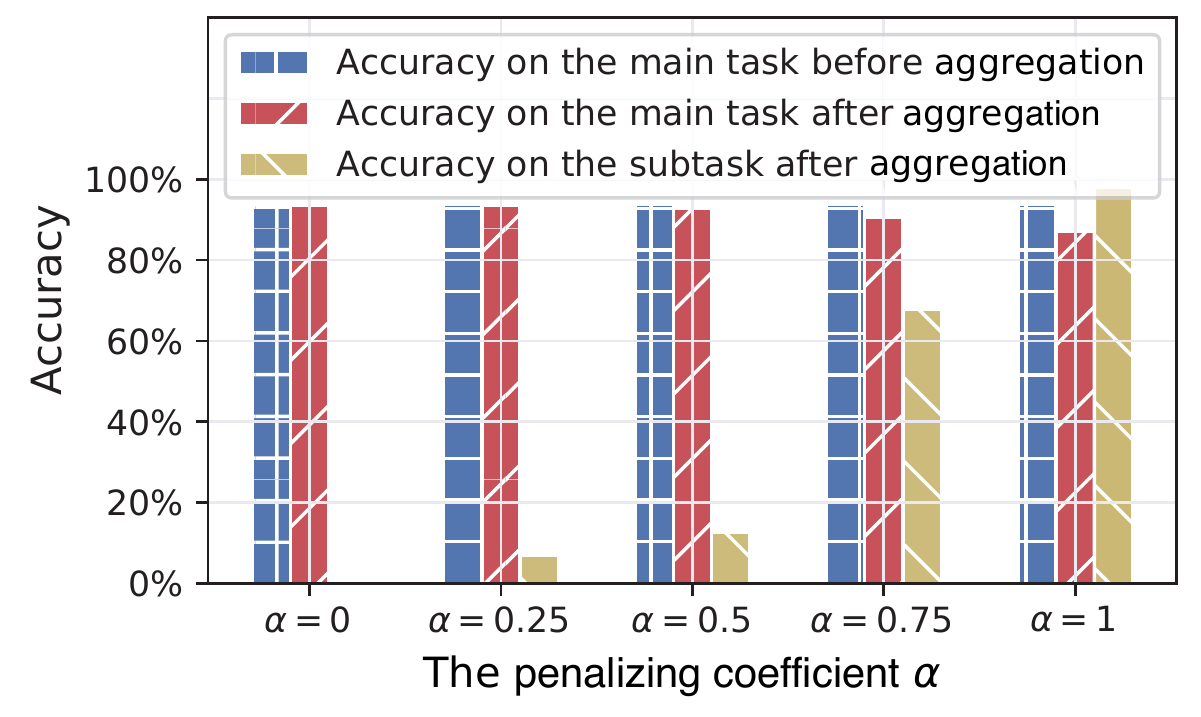}}
  \caption{The accuracy of the global model vs. the penalized coefficient}\label{fig:weight}
\end{minipage}
%\vspace{-3mm}
\end{figure*}

\textbf{The performance of the global model. }In our scheme, determining if an update is anomalous depends on its classification accuracy. If the accuracy is much lower than the current global model, it will be considered as a poisoned update. Therefore, the accuracy of the global model on the main task is a crucial factor to correctly detect anomalous updates. Besides, it is also important for successfully completing the attack, since the backdoor will be forgotten quickly if it is injected when the global model has not converged~\cite{bagdasaryan2018backdoor}.

Figure~\ref{fig:iteration} shows the success rate of attacks at different stages of convergence. For both of the two tasks, we set the number of honest clients as 10 and the malicious client as 1. We can see that even if the attack is successful as expected, it will be forgotten after 20 rounds when the accuracy of the main task is less than 90\%. The prediction results output by a model with low accuracy  are unreliable, and the attack will fail when the global model has acceptable performance. Based on these results, we set the accuracy of the main task to be 90\% in the following experiments.

\textbf{Evaluating the penalizing coefficients.} When a sub-model is considered to be anomalous by some clients, it is penalized by the coefficient in aggregation rather than excluded. Figure~\ref{fig:weight} shows the effect of the penalizing coefficient on the aggregation. We assume that there are 10 sub-models and one of them completes the attack.

If no one reports the results about the poisoned sub-model, the attack will be considered as successful. If anyone reports it and sets the penalizing coefficient $\alpha = 0.75$, the attack still cannot be mitigated. When the poisoned sub-model is penalized by 0.5, the accuracy of the subtask is about 15\%, which means that the attack has a low success rate. Therefore, we conclude that it is appropriate to set the initial value of the coefficient to be 0.5.
Even a sub-model is misclassified by a client, it still can help to improve the accuracy of the global model. And two reports on the same poisoned update reduce the success rate of attack to 10\%. These results confirm that setting the coefficient as 0.5 is suitable.

\textbf{Client-level privacy.} To achieve client-level privacy, the effectiveness of detection may be degraded since the clipping bounds and added noises will reduce the influence of the poisoned update. Figure~\ref{fig:dp} shows the effects of record-level privacy on the detection.
%The settings are the same as Figure~\ref{} and~\ref{}, and
We vary the variance of noises from 0.001 to 0.01 and set the clipping bound to be 15 as did in~\cite{mcmahan2017learning}. Compared with the previous results which do not include privacy protection, we find that client-level privacy does not cause significant accuracy loss when the variance is smaller than 0.003. Therefore, the detection rate based on auditing the model's accuracy will not be affected by the noises.

\textbf{Computation and communication costs. }The main limitation of our scheme lies in the computation costs for testing the updates on the client side and the communication costs of transmitting sub-models to clients.

On the client side, since reporting the results is very efficient, the main cost is derived from downloading the sub-models. During the detection, the communication cost increases linearly with the number of sub-models. If the number of clients is much larger than the number of sub-models, evaluating two models for each client is enough for successful detection.
On the server side, the communication cost is negligible since the number of sub-models is much less than the number of clients.

In our experimental evaluations, we set the total number of clients to be 100, and in each round the server chooses 50 clients to update the model and aggregates them into 10 sub-models. Each sub-model is delegated to 3 clients and every client detects 3 sub-models. Therefore, the server chooses 10 clients to perform the detection. The size of ResNet20 (LeNet) is 3.67MB (6.34MB). The communication cost for detection between the server and the clients is 110.1MB (190.2MB), which is practical as far as we concern.

\begin{figure}[t]
\begin{minipage}[t]{1\columnwidth}
\centering
\includegraphics[width=0.98\columnwidth]{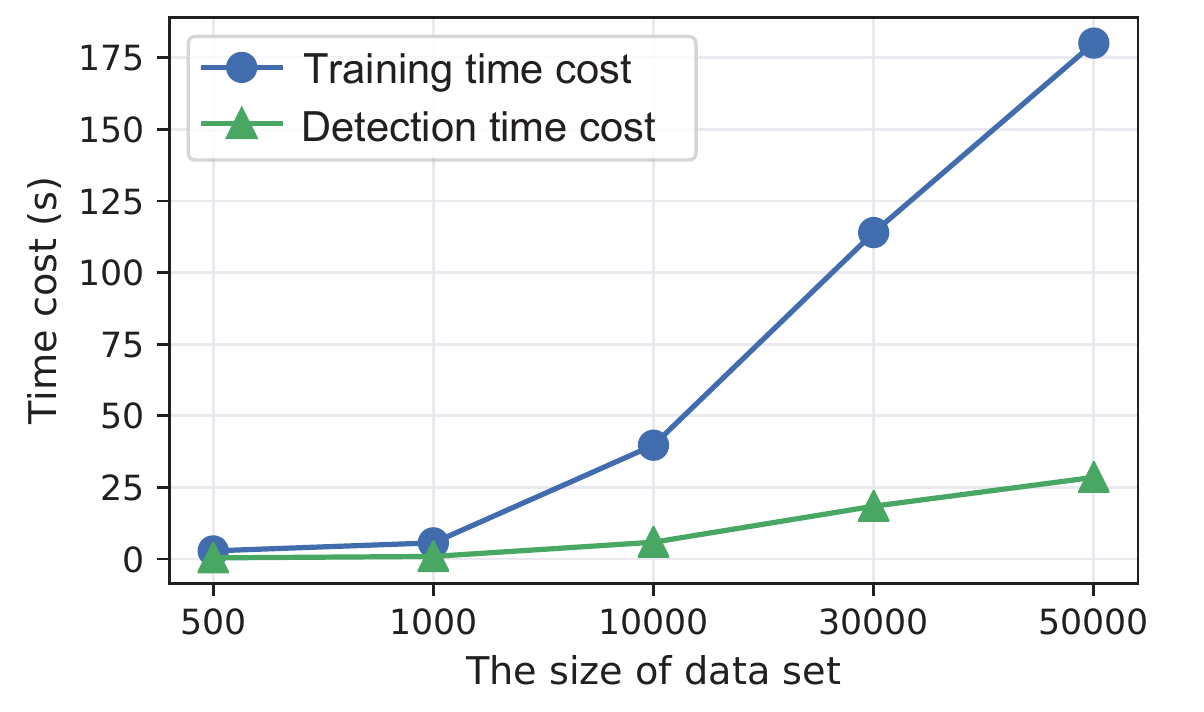}
\caption{Computation costs with different size of the test set}
\label{fig:computation}
\end{minipage}

%\vspace{-3mm}
\end{figure}

The server does not have significant computation cost as the detection is performed on the client side. Obviously, the cost depends on the size of the dataset for each client. Note that testing the accuracy of the received sub-model is much more efficient than training. Figure~\ref{fig:computation} presents the computation costs of detection and training. The results can confirm this observation.

\begin{figure}[t]
\begin{minipage}[t]{0.99\columnwidth}
\centering
\includegraphics[width=0.98\columnwidth, height=5.2cm]{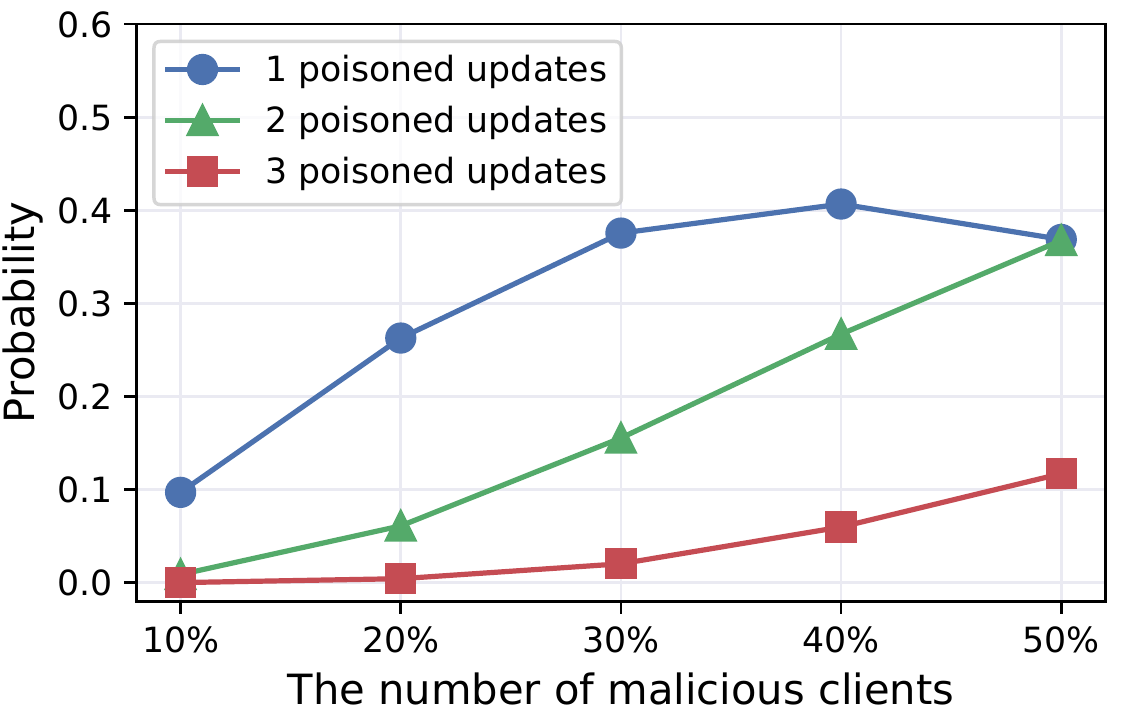}
\caption{The probability of evading the detection vs. the proportion of malicious clients}
\label{fig:probability}
\end{minipage}
\vspace{-4.5mm}
\end{figure}

\textbf{The number of malicious clients.} If there is a large proportion of malicious clients, the returned results from delegated clients may be compromised. Figure~\ref{fig:probability} presents the effect of the proportion of malicious clients on generating a sub-model. We measure the effect by computing the probability that a poisoned sub-model is delegated to other malicious clients. We set that the server chooses 100 clients to update the model in every round, aggregates their updates into 10 sub-models and delegates every sub-model to 3 clients. The probability of tampering the results increases with the increasing of the proportion of malicious clients. Even if only one honest client is reporting the anomaly, the success rate of the attack will be decreased.

\section{Related Work}
\textbf{Poisoning attacks. }The original poisoning attacks try to break the integrity of a machine learning model or a cyber-physical system by changing its behaviors when receiving some specific inputs. Many attacks aim at changing the labels of the training data~\cite{bhagoji2019analyzing}, modifying the original data examples~\cite{gu2017badnets} or injecting  well-designed parameters into the trained model~\cite{dumford2018backdoor, ji2018mod}. Typical examples include evading spam filters, malware detections, and smart grid state estimation~\cite{ZhangSHHLWG19}.

Obviously, the above methods cannot be implemented in collaborative learning. Firstly, the training data are held by multiple clients and cannot be accessed by any other party. Secondly, the training algorithm (i.e., the structure of the model) is defined by the server and known by all the clients. An anomalous model with a special structure will be detected immediately when it is observed by the server or a client.

Although it is difficult for the attacker to control other clients, generating a poisoned model locally is easy to be realized. However, such a trivial attack cannot affect the behavior of the aggregated model. The main reason is that the influence of the poisoned model will be mitigated when the updates are averaged. Therefore, generating sybils and the constrain-and-scale methods are the possible approaches to enhance the poisoned model by controlling multiple clients and scaling up the updates~\cite{shen2016uror, fung2018mitigating, bagdasaryan2018backdoor}.

\begin{figure*}[t]
\centering
  \subfigure[MNIST]{
  \begin{minipage}{0.99\columnwidth}
    \centering
    \includegraphics[width = 0.98\columnwidth]{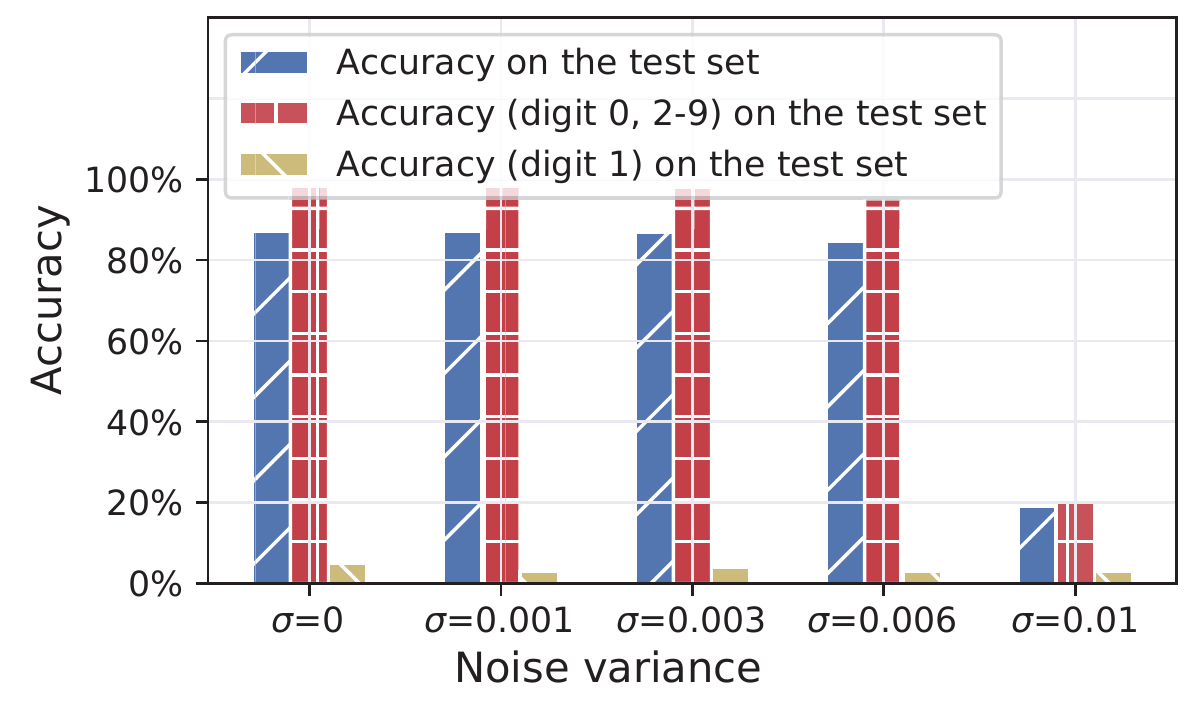}\vspace{5pt}
  \end{minipage}
    }
    \subfigure[CIFAR-10]{
  \begin{minipage}{0.99\columnwidth}
    \centering
    \includegraphics[width = 0.98\columnwidth]{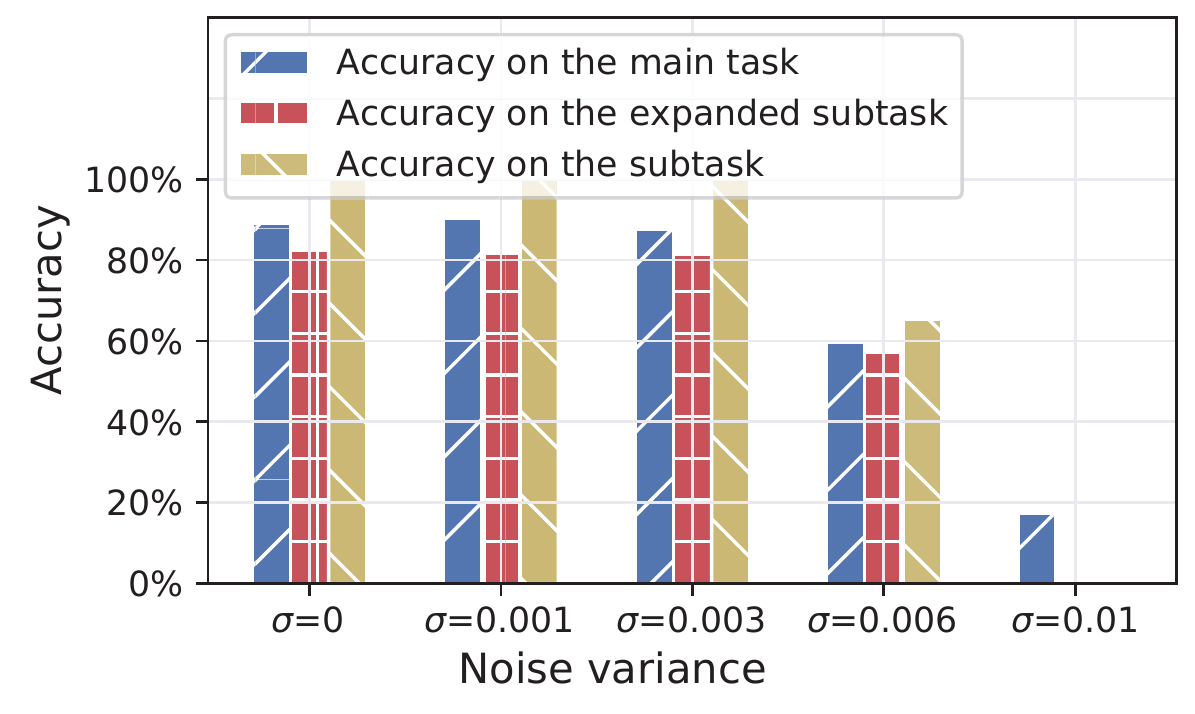}\vspace{5pt}
  \end{minipage}
    }
    \vspace{-3mm}
  \caption{The effect of the Gaussian noise on the accuracy of the sub-model}\label{fig:dp}
\label{fig:dp}
\end{figure*}

Besides the poisoning attacks, there are some other methods that can cause a trained model to misclassify special inputs. Adversarial examples that add noises to the input have been widely investigated recently~\cite{kurakin2016adversarial, huang2017adversarial, tdsc/WangSZZSW19} and proven to be effective to impact real-world systems such as speech recognition~\cite{cm/HuSQLWW19} and autonomous driving~\cite{RenWWQL19}. Moreover, the data preprocessing stage also has flaws~\cite{chen2020scaling}. The main difference between them is that poisoning attacks need to tamper correct labels in the training dataset. Note that it does not mean that poisoning attacks have a stronger assumption about the attacker's ability since the attacker can tamper his/her data arbitrarily in collaborative learning.

\textbf{Defense approaches. }
Some prior works try to detect poisoned model in the training data by evaluating data points with the trained model, utilizing trusted data to train an anomaly detector, pruning the trained model or identifying features associated with the class~\cite{baracaldo2017mitigating, steinhardt2017certified, liu2018fine}. These methods are not applicable in collaborative learning since the server has no access to the training data.

To tackle this problem, new methods have been proposed which only need to observe the updates. Auror~\cite{shen2016uror} provides a statistical mechanism to complete anomaly detection by using k-means algorithm to cluster the updates of clients. However, it assumes that the attack is executed in every training round. Prior results~\cite{bagdasaryan2018backdoor} have shown that a single-round attack is still effective. On the other hand, when the training data of clients are non-IID, it will reduce the accuracy of clustering. Moreover, if the attacker can control 5\% of clients, the attack will achieve 50\% success rate on the main task while evading the detection.

\begin{table*}[t]
    \caption{Comparison with the state-of-the-art solutions}
    \begin{center}
        \begin{tabular}{c|c|c|c|c|c|c|c}
            \Xhline{1.2pt}

			Method & Technique &Single-shot attack  & IID data  & non-IID data  & Single attacker & Sybils & Preventing convergence \\
            \hline
			\textbf{Auror~\cite{shen2016uror}} & Clustering &{$\times$} &{$\checkmark$} & {$\times$} & {$\checkmark$} &  {$\times$}& {$\checkmark$} \\
			\hline
      \textbf{FoolsGlod~\cite{fung2018mitigating}} & Measuring similarities &{$\checkmark$} &{$\checkmark$} & {$\checkmark$} & {$\times$} &  {$\checkmark$}& {$\checkmark$} \\
      \hline
      \textbf{Krum~\cite{blanchard2017machine}} & Minimizing distances &{$\checkmark$} &{$\checkmark$} & {$\times$} & {$\checkmark$} &  {$\checkmark$}& {$\times$} \\
      \hline
			{\textbf{Ours}} & Accuracy auditing &{$\checkmark$} &{$\checkmark$} & {$\checkmark$} & {$\checkmark$} &  {$\checkmark$} & {$\checkmark$} \\
            % \bottomrule
            \Xhline{1.2pt}
        \end{tabular}
        \label{Comparison}

\hfil
    \end{center}
    \label{tab:comparison}
\end{table*}

Evaluating the cosine similarities of updates is a possible defense solution~\cite{fung2018mitigating}. The key idea is that the directions of poisoned updates which are derived from the sybils are similar. If multiple updates are similar to each other, all of them will be discarded when performing the aggregation. This method is especially suitable for the non-IID setting where the distributions of updates may vary significantly with the clients. But it cannot defend against a single malicious client. Besides, the attacker can evade this detection by decomposing the poisoned model into several orthogonal vectors and letting every controlled client update one component~\cite{fung2018mitigating}.

Byzantine tolerant distributed learning is another research area which can eliminate the effect of abnormal updates in collaborative learning while ensuring the convergence~\cite{blanchard2017machine, shayan2018biscotti}. Since these methods assume that the goal of the attacker is to reduce the performance of the trained model, they can not prevent the poisoning attack where the attacker aims to maintain the model's performance on the main task. Moreover, convergence is hard to be preserved when the training data are non-IID. In Table~\ref{tab:comparison}, we have summarized the differences between our scheme with existing solutions.

\textbf{Privacy-preserving machine learning. }
There exists a long line of work which aims to protect the privacy of the sensitive training data. Cryptographic tools such as homomorphic encryption and secure multi-party computation have been leveraged to complete machine learning algorithms on encrypted data~\cite{gilad2016cryptonets, liu2017oblivious, mohassel2017secureml}. These techniques can only be used in a centralized setting, i.e., all the operations are computed on a server (or several non-colluding servers). This makes the training process vulnerable to the poisoning attack. On the other hand, these techniques do not apply to collaborative learning.

Differential privacy is another popular technique which can prevent privacy leakages~\cite{shokri2015privacy, geyer2017differentially, abadi2016deep, mcmahan2017learning, ZhaoWZZC19}. It can be combined with the distributed setting to achieve lower computational costs. Concretely, there are two flavors of the definition of privacy. One is record-level privacy which aims to protect the existence of every record in the training dataset~\cite{abadi2016deep}. The other one called client-level privacy aims to protect the participation of every client in the training process~\cite{geyer2017differentially, mcmahan2017learning}\footnote{In~\cite{mcmahan2017learning}, these two methods are called ``example-level privacy'' and ``user-level privacy''.}. In short, all the updates are clipped to a bound, then random noises are added to the gradients. As all the clients cannot access other clients' datasets and updates, and each client contributes multiple sensitive data when training, protecting the whole dataset of a client is more important in collaborative learning~\cite{geyer2017differentially}. The prior results~\cite{bagdasaryan2018backdoor} show that using differential privacy in collaborative learning can mitigate the poisoning attack as clipping and perturbing the uploaded parameters will change the direction of the poisoned model. However, the attack can still be successful when the attacker controls more clients to upload poisoned updates.

Secure aggregation is another scenario of protecting the privacy of updates in collaborative learning~\cite{bonawitz2017practical}. It can prevent any other party from knowing the specific updates of clients. To this end, the server can no longer use any method to detect anomalies of the updates. Even if there exists a method that can detect the poisoning attack from the trained model, collaborative learning will fail since there is no way to determine the identity of the attacker in secure aggregation.

\section{Conclusion}
In this paper, we investigated the poisoning attack in collaborative learning system and presented a novel scheme for detecting anomalous updates.
By realizing client-side cross-validation, we delegated the detection task to the clients who are able to evaluate the performance of the updates. Our scheme can apply to both IID and non-IID setting. Besides, it can also protect the client-level privacy by integrating differential privacy to our design. The experimental results demonstrated that our scheme is robust to the existing poisoning attacks.

\bibliographystyle{IEEEtran}
\bibliography{detection}

\end{document}